\documentclass[12pt]{article}
\usepackage[totalwidth=470truept,totalheight=636truept]{geometry}
\usepackage{latexsym,graphicx,amsfonts,amssymb,amsmath,color,dsfont,cancel,tikz,multirow,slashed}

\usepackage[numbers, sort&compress]{natbib}

\usepackage[hypertexnames=false,hidelinks]{hyperref}

\DeclareMathAlphabet{\mymathbb}{U}{BOONDOX-ds}{m}{n}
\def\theequation{\arabic{section}.\arabic{equation}}

\renewcommand{\theequation}{\thesection.\arabic{equation}}
\global\arraycolsep=1truept
\renewcommand{\theequation}{\arabic{section}.\arabic{equation}}

\hypersetup{colorlinks=true,
    linkcolor=black,
    filecolor=black, 
    citecolor=black,
    urlcolor=blue}

\RequirePackage{booktabs}

\begin{document}

\null

\vskip1truecm

\begin{center}
{\LARGE \textbf{Massive Higher-Spin Multiplets and}}

\vskip.8truecm

{\LARGE \textbf{Asymptotic Freedom in Quantum Gravity}}

\vskip1truecm

\textsl{{\large Marco Piva}}

\vskip .1truecm

{\textit{National Institute of Chemical Physics and
Biophysics,\\ R\"{a}vala 10, Tallinn 10143, Estonia}}

\vspace{0.2cm}

marco.piva@kbfi.ee
\vskip1truecm

\textbf{Abstract}
\end{center}

We consider massive higher-spin multiplets coupled to quantum gravity and compute their contributions to the gravitational beta functions at one loop. Such theories, if quantized with the Feynman prescription, would contain ghosts. Instead, those degrees of freedom are turned into purely virtual particles by means of the fakeon prescription and the resulting theories are both renormalizable and unitary. We extract the necessary number of fields to obtain asymptotic freedom by solving the relevant renormalization-group equations in the ultraviolet limit. We show that in the case of irreducible rank-$s$ bosons with minimal kinetic term only a certain number of fields with $0\leq s \leq 3$ is compatible with asymptotic freedom. No solution involving fermionic rank-$s$ fields are found. We also consider the cases of rank-$1$ and symmetric rank-$2$ tensor fields with nonminimal kinetic terms and repeat the analysis, finding that the allowed number of fields is enlarged. Possible implications for metric affine theories of gravity are pointed out.

\vfill\eject

\section{Introduction}
Asymptotic freedom is a property of certain quantum field theories which ensures that their interactions become weaker at high energies. Theories with such a feature belong to a special class in which perturbative techniques can be trusted up to arbitrary energy scales. 

In particle physics the most relevant theory in this class is quantum chromodynamics. This follows from the general result that any non-Abelian gauge theory is asymptotically free~\cite{Gross:1973id,Politzer:1973fx}. On the other hand, quantum electrodynamics does not belong to this class of theories. For this reason, the standard model is not asymptotically free. 

In the case of gravitational interactions we have a similar situation. A renormalizable extension of Einstein gravity is Stelle theory~\cite{Stelle:1976gc}, which is plagued by the presence of unstable ghosts that violate unitarity. This problem is solved by turning the ghosts into purely virtual degrees of freedom (or ``fakeons") with the help of a different quantization prescription~\cite{Anselmi:2017ygm,Anselmi:2017yux}. The new theory is physically different from the Stelle one~\cite{Anselmi:2018tmf} but shares the same properties under renormalization~\cite{Anselmi:2018ibi}. In particular, asymptotic freedom cannot be obtained unless one of the degrees of freedom is tachyonic~\cite{Avramidi:1985ki}. Since the fakeon prescription cannot be applied to tachyons, that option is not viable. The result is a renormalizable and unitary theory where some couplings tend to zero in the ultraviolet (UV) while others do not, pretty much like in the standard model. The coupling to standard matter fields does not improve the situation~\cite{Fradkin:1981iu}. 

Another possibility, investigated in the literature, is to introduce additional higher-derivative terms with dimension greater than four and of the Lee-Wick type~\cite{Modesto:2015ozb,Modesto:2016ofr,Anselmi:2017ygm}. Those theories are superrenormalizable but, if quantized using the Feynman prescription, contain pairs of complex conjugate ghost-poles that violate unitarity, Hermiticity and the locality of counterterms~\cite{Aglietti:2016pwz}. The ghosts can be quantized with the fakeon prescription, removing all the inconsistencies. Moreover, operators at least cubic in the curvature can be introduced, together with renormalization-group (RG) chains~\cite{Anselmi:2017ygm}, in order to make this class of theories finite. The RG equations in the six-derivative case have been recently studied in~\cite{Rachwal:2021bgb}. Although superrenormalizable theories of gravity can be made consistent by means of the fakeon prescription, they are infinitely many, raising the problem of uniqueness.

In this paper we consider the only strictly renormalizable theory of quantum gravity and show that it can be made asymptotically free by coupling it to other fakeons that might exist in nature. In particular, we consider the massive higher-spin multiplets of ref.~\cite{Anselmi:2020opi}, where each irreducible rank-$s$ multiplet contains both standard particles and fakeons. To our knowledge this is the only way to couple interacting matter field with spin higher than 1 to quantum gravity without violating either renormalizability or unitarity. Indeed, if we quantize a whole multiplet with the Feynman prescription we would introduce several ghosts, instead of fakeons. On the other hand, it is possible to eliminate the ghost degrees of freedom by imposing certain symmetries for the kinetic terms. However, this would define unitary but nonrenomalizable theories~\cite{Singh:1974qz,Singh:1974rc} (once the interactions are included), analogous to the Proca~\cite{Proca:1936fbw}, Pauli-Fierz~\cite{Fierz:1939ix} and Rarita-Schwinger~\cite{Rarita:1941mf} ones, which can be viewed as effective field theories~\cite{Bellazzini:2019bzh} but not as fundamental ones.

A crucial difference between the massive multiplets of this paper and more standard ones is that the former can be coupled to a general background without introducing any additional ghost. For example, in Pauli-Fierz theory the coupling to gravity turns on the physical propagation of the (ghostly) trace of the field, known as Boulware-Deser ghost~\cite{Boulware:1972yco}, which is non-dynamical in the flat-space limit. This is not the case in our approach since the fields are traceless and all the degrees of freedom contained in the higher-spin multiplets are already present in flat spacetime. In particular, all the potential ghosts are those associated to the poles with negative residues, which are turned into purely virtual particles by means of the fakeon prescription. No Boulware-Deser ghost is turned on and the theory can be safely coupled to any curved background. Moreover, general reducible multiplets with nonvanishing trace can be considered along the same line. Also in that case there is no room for the Boulware-Deser ghost to arise. In fact, the trace of the field would be already present expanding around flat spacetime, since we always consider the most general kinetic and mass terms. Then, all the would-be ghosts are prescribed with the fakeon prescription. 

Alternative ways of getting rid of the Boulware-Deser ghost are known in the literature, such as the de Rham-Gabadadze-Tolley model~\cite{deRham:2010kj,deRham:2010ik,Hassan:2011hr} and bimetric theories~\cite{Hassan:2011zd}. However, in those cases renormalizability is lost.

Massive higher-spin fields are also considered in non-conventional representations of the Lorentz group and used to directly derive $S$-matrix elements without relying on a Lagrangian description~\cite{Weinberg:1964cn}. This framework has been recently adopted in the context of dark matter phenomenology~\cite{Criado:2020jkp,Criado:2021itq}.

Finally, higher-spin gauge theories in four dimensions can be consistently formulated in the free-field limit and in flat spacetime~\cite{Fronsdal:1978rb,Fang:1978wz} or in anti-de~Sitter spacetime, while the Lagrangian formulation of their self-interacting version is still subject of investigation~\cite{Vasiliev:1990vu,Bekaert:2010hw}. In general, these theories can be coupled to external gravity only if the spacetime satisfies certain consistency conditions~\cite{Christensen:1978md}.  However, these issues do not appear in the special multiplets considered in this paper. In fact, we do not consider massless theories, since fakeons can only be massive, in order to avoid violation of causality at arbitrary distances~\cite{Anselmi:2018tmf,Anselmi:2019nie}.

The paper is organized as follows. In~\autoref{fakeonpres}, we introduce the details of the fakeon prescription and the related projection. In~\autoref{sec:betaQG}, we briefly review the renormalization properties of quantum gravity. In~\autoref{sec:bosons}, we consider the bosonic higher-spin fields with minimal kinetic term coupled to quantum gravity and compute their contribution to the gravitational beta functions at one loop. In~\autoref{sec:fermions}, we repeat the same steps of~\autoref{sec:bosons} in the case of fermionic higher-spin multiplets. In~\autoref{AF}, we derive and solve the equations necessary to obtain asymptotic freedom. In~\autoref{examples}, we repeat the steps of~\autoref{sec:bosons} and~\autoref{AF} in the case of rank-1 and symmetric rank-2 multiplets with nonminimal kinetic terms. Some special cases are presented in details. In~\autoref{sec:mag}, we comment on the possibility that metric affine theories of gravity might be asymptotically free and compute the contribution of hook-(anti)symmetric rank-3 tensors to the beta functions. Section \ref{sec:concl} contains our conclusions. \autoref{lorentz} collects the formulas of Lorentz algebra used in the paper. \autoref{projectors} collects the definitions of the spin-2 projectors.

We use the signature $(+,-,-,-)$ for the metric tensor. The Riemann and Ricci tensors are defined as $R^{\mu}_{ \ \nu\rho\sigma}=\partial_{\rho}\Gamma^{\mu}_{\nu\sigma}-\partial_{\sigma}\Gamma^{\mu}_{\nu\rho}+\Gamma^{\mu}_{\alpha\rho}\Gamma^{\alpha}_{\nu\sigma}-\Gamma^{\mu}_{\alpha\sigma}\Gamma^{\alpha}_{\nu\rho}$ and $R_{\mu\nu}=R^{\rho}_{ \ \mu\rho\nu}$, respectively. We write the four-dimensional integrals over spacetime points of a function $F$ of a field $\phi(x)$ as $\int\sqrt{-g}F(\phi)\equiv\int\text{d}^4x\sqrt{-g(x)} F\left(\phi(x)\right)$. The $D$-dimensional integrals are distinguished from the four-dimensional ones by the presence of the factor $\mu^{-\varepsilon}$. We always assume that the integral of the Gauss-Bonnet term vanish, i.e. $\int\sqrt{-g}\left(R_{\mu\nu\rho\sigma}R^{\mu\nu\rho\sigma}-4R_{\mu\nu}R^{\mu\nu}+R^2\right)=0.$

\section{Fakeon prescription/projection}
\label{fakeonpres}
The fakeon is a degree of freedom of a new type which can only be a virtual state. In quantum field theory this is achieved in two steps: a prescription for the propagator, to mantain the fakeon as a possible virtual state, and a projection, to remove it from the physical spectrum. This procedure ensures the unitarity of the projected theory without losing the properties under renormalization of the unprojected one. More explicitly, the fakeon propagator reads
\begin{equation}\label{fakeon}
\lim_{\mathcal{E}\rightarrow 0}i\frac{p^2-m^2}{(p^2-m^2)^2+\mathcal{E}^4}.
\end{equation}
The distribution defined by the fakeon prescription is not the Cauchy principal value, since the loop integrals must be performed along special deformed domains. Using~\eqref{fakeon} without any prescription for the integration domains is called ``Feynman-Wheeler propagator" which, if used inside loop diagrams, leads to violation of stability, unitarity and the locality of counterterms~\cite{Anselmi:2020tqo}. 

The fakeon prescription is equivalent to treat the amplitudes associated to fake processes by means of a nonanalytic Wick rotation, defined by the so-called ``average continuation"\cite{Anselmi:2017yux}
\begin{equation}\label{av}
\mathcal{A}_{\text{f}}=\frac{1}{2}\left(\mathcal{A}_++\mathcal{A}_-\right),
\end{equation}
where $\mathcal{A}_+$ and $\mathcal{A}_-$ are the analytic continuations from above and below the cut associated to the fakeon threshold, respectively. The Minkowski amplitudes for standard particles, instead, are given by $\mathcal{A}_+$, which is obtained by means of the usual Wick rotation. In the case of amplitudes with multiple thresholds and mixed contributions from both particles and fakeons, each threshold must be treated accordingly (see~\cite{Anselmi:2021hab} for the general rules).

At the level of the Fock space W, the fakeon prescription works as follows. Starting from the unitarity condition for the $S$-matrix, written as $S=\mathds{1}+iT$, we obtain
\begin{equation}
-i(T-T^{\dagger})=TT^{\dagger},
\end{equation}
which is known as the ``optical theorem". In terms of matrix elements
\begin{equation}\label{opt}
\langle \alpha|(T-T^{\dagger})|\beta\rangle=i\sum_{|n\rangle}\langle\alpha|T|n\rangle\langle n|T^{\dagger}|\beta\rangle, \qquad \sum_{|n\rangle}|n\rangle \langle n|=\mathds{1}, \qquad |\alpha\rangle, |\beta\rangle, |n\rangle \in \text{W}.
\end{equation}
Moreover, any quantum field theory satisfies a set of diagrammatical identities, called ``cutting equations'', that can be recollected in the form of a pseudo-unitarity equation
\begin{equation}\label{popt}
\langle \alpha|(T-T^{\dagger})|\beta\rangle=i\sum_{|n\rangle}\langle\alpha|T|n\rangle\sigma_n \langle n|T^{\dagger}|\beta\rangle,
\end{equation}
where, for a fixed diagram, $\sigma_n=\pm 1$ depending on the sign of the residues of the propagators involved. A unitary theory has $\sigma_n=1$ for every diagram, while a theory which contains a ghost has $\sigma_n=-1$ for a subset of diagrams, leading to the violation of unitarity.

Suppose we want to turn a certain number of degrees of freedom into fakeons. Let $\bar{\text{V}}$ be the subspace of the states generated by applying to the vacuum state at least one creation operator of the fields associated to those degrees of freedom and V be its complimentary. Assuming that the propagators of the degrees of freedom in V have positive residue, equation \eqref{popt} can be split into

\begin{equation}\label{splitpopt}
\langle \alpha|(T-T^{\dagger})|\beta\rangle=i\sum_{|m\rangle\in \text{V}}\langle \alpha|T|m\rangle\langle m|T^{\dagger}|\beta\rangle+i\sum_{|\bar{m}\rangle\in \bar{\text{V}}}\langle \alpha|T|\bar{m}\rangle\sigma_{\bar{m}}\langle \bar{m}|T^{\dagger}|\beta\rangle.
\end{equation}
The fakeon prescription sets $\sigma_{\bar{m}}=0$. Finally, we can consistently project the external states onto V and obtain 
\begin{equation}\label{subopt}
\langle a|(T-T^{\dagger})|b\rangle=i\sum_{|m\rangle\in \text{V}}\langle a|T|m\rangle\langle m|T^{\dagger}|b\rangle, \qquad |a\rangle , |b\rangle \in \text{V},
\end{equation}
which is the unitarity equation in the subspace V, i.e. the degrees of freedom propagating through the cuts are the same as the external ones.

The fakeon degrees of freedom are removed also at classical level. In fact, the projection survives the classical limit. They could be seen as ``auxiliary fields with a kinetic term". Typically, auxiliary fields do not have a kinetic term and they can be easily integrated out. If a kinetic term is present, then at classical level the initial conditions need to be specified. In the case of fakeons the initial conditions are ``frozen" by the prescription, i.e. the Green functions are fixed and there is no freedom in the choice of initial conditions.  The true classical (nonlocal) action, where all the fakeons are integrated out according to the prescription, cannot be derived without the knowledge of the nonperturbative fakeon projection. Therefore, it is convenient to use a local ``interim" classical action where we can formulate Feynman rules and make computations by applying the procedure explained above. Moreover, perturbative techniques can be applied to obtain the projected classical action as an expansion~\cite{Anselmi:2018bra}.

We remark that in principle different prescriptions compatible with unitarity might exist~\cite{Anselmi:2017yux}. However, the fakeon prescription allows us to keep the renormalization properties of the theory untouched. This is the key feature for the formulation of quantum gravity, which is the main application of the idea of purely virtual particles. Indeed, the strictly renormalizable theory of gravitational interactions~\cite{Stelle:1976gc} propagates ghost degrees of freedom, which is physically unacceptable. The presence of ghosts spoils the unitarity of theory, even in the cases where those degrees of freedom are unstable, since there are situations where they can be considered long lived. The fakeon prescription allows to turn the ghost degrees of freedom into purely virtual particles. In this way the resulting theory is both renormalizable and unitary (in the subspace where fakeons are projected away). Moreover, since the prescription changes the definition of the amplitudes that involve fakeon thresholds, the theory gives predictions that are physically different from those of the theory with ghosts. 

In the case of massive higher-spin theories we are in an analogous situation. They are renormalizable but propagate multiplets of particles with different spin that contain both ghosts and standard particles. It is possible to apply the fakeon prescription to turn the ghosts of massive higher-spin theories into purely virtual ones and obtain renormalizable and unitary theories \cite{Anselmi:2020opi}. In the next sections we couple those theories to quantum gravity and fix the number of higher-spin fields necessary to obtain asymptotic freedom in the gravitational coupling constants.

Finally, we remind that any non-tachyonic degree of freedom can be turned into a fakeon by means of the procedure described above, even standard particles. The freedom in the choice of the prescription is acceptable as long as the theory is consistent and produces predictions that in principle could be tested. In this way, theories quantized with different choices for prescriptions can be discriminated from one another and potential future experiments may determine whether purely virtual particles exist in nature. This possibility has been recently explored in the context of particle physics phenomenology~\cite{Anselmi:2021icc,Anselmi:2021chp}.   

\section{Beta functions in quantum gravity}
\label{sec:betaQG}
\setcounter{equation}{0}
The renormalizable and unitary theory of quantum gravity with fakeons is defined by the classical action 
\begin{equation}
S_{\text{QG}}(g)=-\frac{1}{2\kappa ^{2}}\int \sqrt{-g}\left[
2\Lambda+\zeta R+\frac{1}{2\alpha}C^2-\frac{1}{6\xi}R^{2}\right] ,  \label{lhd}
\end{equation}%
where $\alpha $, $\xi $, $\zeta $, $\Lambda$ and $\kappa $ are real
constants, with $\zeta >0$, $\alpha >0$, $\xi >0$ (this last choice is explained below) and $C^2\equiv C_{\mu\nu\rho\sigma}C^{\mu\nu\rho\sigma}$ is the square of the Weyl tensor. The theory contains a masselss spin-2 graviton, a massive scalar and a massive spin 2. The latter must be quantized with the fakeon prescription in order to have both renormalizability and unitarity. The scalar can be quantized in either way. Remarkably, if quantized with the Feynman prescription, it can be viewed as the inflaton and used to study primordial cosmology~\cite{Anselmi:2020lpp, Anselmi:2020shx}. This leads to predictions that can be discriminated from other models of inflation and possibly be tested in the near future~\cite{Hazumi:2019lys}.

We stress that the action~\eqref{lhd} is the interim, unprojected one. Therefore, the true classical action is obtained by first following the procedure described in~\autoref{fakeonpres}, then project the spin-2 fakeon away from the physical spectrum and finally take the classical limit.

Since the fakeon prescription does not affect the counterterms, the beta functions of the theory~\eqref{lhd} are the same as in Stelle theory. In particular, the counterterms can be written as\footnote{This result can be obtained in the Batalin-Vilkovisky formalism up to a cohomologically exact term that is removed by a canonical transformation. For more details see~\cite{Anselmi:2015niw,Anselmi:2018ibi}}
\begin{equation}
S_{\text{QG}}^{\text{ct}}=\frac{\mu ^{-\varepsilon }}{(4\pi )^{2}\varepsilon }\int 
\sqrt{-g}\left[ 2\Delta \Lambda+\Delta \zeta R+\frac{\Delta \alpha}{2}C^2 -\frac{\Delta \xi }{6}R^{2}%
\right],  \label{scount}
\end{equation}%
where $\Delta \Lambda$, $\Delta \zeta $, $\Delta \alpha $ and $\Delta
\xi $ are constants, $\mu $ is the
dynamical scale and $\varepsilon =4-D$, $D$ being the continued spacetime dimension introduced by the dimensional regularization.

The beta functions are extracted from \eqref{scount} and read
\begin{equation}
\beta _{\alpha }=\alpha^2\Delta \alpha ,\qquad
\beta _{\xi }=\xi^2\Delta \xi ,\qquad \beta
_{\zeta }=-\Delta \zeta ,\qquad \beta
_{\Lambda}=-\Delta \Lambda,
\label{betas}
\end{equation}
where the beta function $\beta_{\lambda}$ of a coupling constant $\lambda$ at a scale $\mu_0$ is defined as
\begin{equation}
    \beta_{\lambda}=\frac{\text{d}\lambda}{\text{d}t}, \qquad t\equiv\frac{2\kappa ^{2}}{(4\pi )^{2}}\ln\left(\frac{\mu_0}{\mu}\right).
\end{equation}
The coefficients $\Delta \Lambda$, $\Delta \zeta $, $\Delta \alpha $
and $\Delta \xi $ can be worked out by computing the
graviton self energy and the renormalization of the BRST operators. In the case of pure gravity the results are \cite{Avramidi:1985ki,Salvio:2017qkx, Anselmi:2018ibi}\footnote{The formulas in (\ref{deltas}) are related to those in \cite{Anselmi:2018ibi} through the substitutions $\alpha\rightarrow1/\alpha$, $\xi\rightarrow1/\xi$.}
\begin{eqnarray}
\Delta \alpha &=&-\frac{133}{10},\qquad \Delta \xi =\frac{5}{6}+\frac{5\alpha}{%
\xi}+\frac{5\alpha^{2}}{3\xi^{2}},\qquad \Delta \zeta =\zeta \left( 
\frac{5\xi}{6 }+\frac{5\alpha^2 }{3\xi ^{2}}+A\right) ,  \notag \\
\Delta \Lambda &=&\Lambda\left(-5\alpha+2\xi
+2A\right)-\frac{5\zeta^{2}\alpha^2}{4}-\frac{\zeta ^{2}\xi^2}{4},
\label{deltas}
\end{eqnarray}%
where $A$ is a gauge-dependent arbitrary constant. 

In order to have asymptotic freedom in $\alpha$ and $\xi$ a necessary condition is that their beta functions be negative. Using (\ref{deltas}) we have
\begin{equation}
    \beta_{\alpha}=-\frac{133}{10}\alpha^2, \qquad \beta_{\xi}=\frac{5}{6}\xi^2\left(1+6\frac{\alpha}{\xi}+2\frac{\alpha^2}{\xi^2}\right).
\end{equation}
Therefore, when matter is switched off, $\beta_{\alpha}$ is negative and quadratic in $\alpha$, while $\beta_{\xi}$ is polynomial in $\alpha$ and $\xi$ and it can be made negative by changing the sign of one of the parameters. If we do so we would introduce tachyonic degrees of freedom, once we expand the metric around flat spacetime \cite{Avramidi:1985ki}. Tachyons cannot be treated with the fakeon prescription and therefore we need to exclude their presence by imposing $\alpha>0, \xi>0$. We refer to this as the ``no-tachyon condition", which needs to be imposed for every degree of freedom. For this reason $\beta_{\xi}$ is always positive in pure gravity. If we couple standard matter (with zero nonminimal couplings) the contributions are
\begin{equation}
    \Delta\alpha_{\text{SM}}=-\frac{1}{60}\left(N_0+6N_f+12N_{v}\right), \qquad \Delta\xi_{\text{SM}}=\frac{N_0}{12},
\end{equation}
where $N_0$ is the number of scalars, $N_f$ is the number of Dirac fermions plus one half the number of Weyl fermions and $N_v$ is the number of gauge vectors ($N_0=4$, $N_f=45/2$ and $N_v=12$ in the case of the standard model). We see that the signs of the beta functions remain the same. It is possible to modify $\Delta\xi_{\text{SM}}$ by adding nonminimal (and nonconformal) interactions between matter and gravity. For example, if we introduce the nonminimal term $\frac{1}{12}\sqrt{-g}(1+2\eta_0)R\varphi^2$
for every scalar field $\varphi$, then we have
\begin{equation}
    \Delta\xi_{\text{SM}}=\frac{N_0\eta_0^2}{3},
\end{equation}
which cannot change the sign of $\beta_{\xi}$. Additional nonminimal terms can be added for spin higher than 0. Those terms might help in having the correct signs for asymptotic freedom. However, they would break renormalizability unless all the degrees of freedom contained in the higher-spin multiplets are included. This forces us to quantize some of them as fakeons to preserve unitarity, leading to the special multiplets studied in this paper. For this reason, we concentrate only on their minimal coupling to gravity, although in the next section we write the general form of nonminimal terms and show how to include them in the computations.

The coupling to matter generates gravitational one-loop counterterms of the form
\begin{equation}\label{eq:ctmatter}
    S_s^{\text{ct}}(g)=-\frac{\mu^{-\varepsilon}}{(4 \pi)^2\varepsilon}\int\sqrt{-g}\left[\lambda+d R+c \ C^2+bR^2\right].
\end{equation}
The contributions to the beta functions can be read from~\eqref{eq:ctmatter} by means of the relations
\begin{equation}\label{deltamatt}
    \Delta\alpha_{s}=-2c, \qquad \Delta\xi_{s}=6b,\qquad \Delta\zeta_{s}=d, \qquad \Delta\Lambda_{s}=\frac{\lambda}{2},
\end{equation}
where $c$ is called ``central charge" in conformal field theories.

In the next sections, we compute the coefficients~\eqref{deltamatt} for the higher-spin multiplets and study asymptotic freedom.

\section{Bosons}
\label{sec:bosons}
\setcounter{equation}{0}
We start by considering a bosonic rank-$s$ tensor field $\varphi_{\mu_1\dots \mu_s}$, completely symmetric and traceless. In flat spacetime, the most general quadratic action we can build is 
\begin{equation}\label{hsflat}
    S_s(\varphi)=\frac{(-1)^s}{2}\int\left(\partial_{\mu}\varphi_{\nu_1\dots\nu_s}\partial^{\mu}\varphi^{\nu_1\dots\nu_s}+\lambda_s\partial_{\mu}\varphi_{\nu_1\dots\nu_s}\partial^{\nu_1}\varphi^{\mu\nu_2\dots\nu_s}-m_s^2\varphi_{\mu_1\dots\mu_s}\varphi^{\mu_1\dots\mu_s}\right),
\end{equation}
where $\lambda_s$, $m_s^2$ are real constants and the indices are raised and lowered by means of the flat-spacetime metric $\eta_{\mu\nu}=\text{diag}\left(1,-1,-1,-1\right)$. The field $\varphi_{\mu_1\dots \mu_s}$ transforms according to the irreducible representation $\left(\frac{s}{2},\frac{s}{2}\right)$ of the Lorentz group. Therefore, it contains $(s+1)^2$ degrees of freedom, whose sign of the residues at the poles of each propagator are alternating \cite{Anselmi:2020opi}.  For example, with $s=1$ the action
\begin{equation}\label{spin1flat}
    S_1(\varphi)=-\frac{1}{2}\int\left(\partial_{\mu}\varphi_{\nu}\partial^{\mu}\varphi^{\nu}+\lambda_1\partial_{\mu}\varphi_{\nu}\partial^{\nu}\varphi^{\mu}-m_1^2\varphi_{\mu}\varphi^{\mu}\right),
\end{equation}
describes a vector multiplet with mass squared $m_1^2$ and residue $1$ and a scalar with mass squared $m_1^2/(\lambda_1+1)$ and residue $-1/(\lambda_1+1)$ (see~\autoref{examples} for details). Therefore, if we impose the no-tachyon condition $\lambda_1>-1$ the two residues have necessarily opposite signs. This occurs for every multiplets (see \autoref{examples} for $s=2$ and \autoref{sec:mag} for $s=3$). The different sign in the residues ensures the renormalizability of the theory (once renormalizable interactions are switch on), since nonrenormalizable behaviors mutually cancel.

The action \eqref{hsflat} can be coupled to gravity. The minimal quadratic action is 
\begin{equation}\label{hsgeneral}
    S_s(\varphi,g)=\frac{(-1)^s}{2}\int \sqrt{-g}\left(\nabla_{\mu}\varphi_{\nu_1\dots\nu_s}\nabla^{\mu}\varphi^{\nu_1\dots\nu_s}+\lambda_s\nabla_{\mu}\varphi_{\nu_1\dots\nu_s}\nabla^{\nu_1}\varphi^{\mu\nu_2\dots\nu_s}-m_s^2\varphi_{\mu_1\dots\mu_s}\varphi^{\mu_1\dots\mu_s}\right),
\end{equation}
where $\nabla$ is the covariant derivative, and the indices are raised and lowered by means of the metric $g_{\mu\nu}$. Moreover, we can add nonminimal terms
\begin{equation}\label{nonminimal}
    S_s^{\text{nm}}(\varphi,g)=\frac{(-1)^{s+1}}{2}\int\sqrt{-g}R_{\alpha\beta\rho\sigma}\varphi_{\mu_1\dots\mu_s}\varphi_{\nu_1\dots\nu_s}\sum_nc_{sn}\text{G}_n^{\alpha\beta\rho\sigma\mu_1\dots\mu_s\nu_1\dots\nu_s},
\end{equation}
where the sum is over all the independent ways of contracting the Riemann tensor with two $\varphi$ fields, $c_{sn}$ are real constants and $\text{G}_n$ are monomials in the metric tensor. Taking into account the symmetries of the fields $\varphi$ and of the Riemann tensor, the nonminimal terms~\eqref{nonminimal} turns into
\begin{equation}\label{hsnonminimal}
 S_s^{\text{nm}}(\varphi,g)=\frac{(-1)^{s+1}}{2}\int\sqrt{-g}\varphi_{\mu_1\dots\mu_s}\text{B}_s^{\mu_1\dots\mu_s\nu_1\dots\nu_s}\varphi_{\nu_1\dots\nu_s},
\end{equation}
where
\begin{align}\label{bnm}
&\text{B}_0=\eta_{01}R,\qquad\text{B}_1^{\mu\nu}=\eta_{11}Rg^{\mu\nu}+\eta_{12}R^{\mu\nu},\\
&\text{B}_{s>1}^{\mu_1\dots\mu_s\nu_1\dots\nu_s}=\eta_{s1}R g^{\mu_1\nu_1}\dots g^{\mu_s\nu_s}+\eta_{s2}R^{\mu_1\nu_1}g^{\mu_2\nu_2}\dots g^{\mu_s\nu_s}+\eta_{s3}R^{\mu_1\mu_2\nu_1\nu_2}g^{\mu_3\nu_3}\dots g^{\mu_s\nu_s}
\end{align}
and $\eta_{sn}$ are real coefficients. Note that we could have added an additional term in~\eqref{hsgeneral} where $\nabla_{\mu}$ and $\nabla^{\nu_1}$ are switched. However, this would be equivalent to redefine the nonminimal couplings~\eqref{bnm}, since the commutator of two covariant derivatives produces terms proportional to the curvature. Therefore, those terms can be ignored without loss of generality.

Finally, we mention that in principle more general nonminimal interactions can be build when many $\varphi$ fields are present. Schematically, they can be written as 
\begin{equation}
    \int\sqrt{-g}\left[\varpi_{ab}^{s}\mathcal{R}\varphi_{s}^{(a)}\varphi_{s}^{(b)}+\varrho_{s s^{\prime}}\mathcal{R}\varphi_{s}\varphi_{s^{\prime}}\right], \qquad\varpi_{ab}^{s},\varrho_{s s^{\prime}}\in\mathbb{R},
\end{equation}
where the first term represents all the possible contractions between the Riemann tensor and two different $\varphi$ fields with the same spin, while the second term represents all the possible contractions between the Riemann tensor and two $\varphi$ fields with different spin. We do not consider these nonminimal terms in this paper, as well as kinetic terms that mix two different fields $\varphi$.

\subsection[Counterterms]{Counterterms for any spin $s$ with $\lambda_s=0$}\label{counterterms0}

The gravitational one-loop counterterms can be derived in the case $\lambda_{s}=0$ for any spin $s$ by using the algorithm \cite{tHooft:1974toh}, which works as follows.
Consider an action of the form
\begin{equation}\label{thooftact}
    S(\varphi,g)=\int \sqrt{-g}\left[\frac{1}{2}\partial_{\mu}\varphi_i\delta^{ij}g^{\mu\nu}\partial_{\nu}\varphi_j+\varphi_i\left(\text{N}^{\mu}\right)^{ij}\partial_{\mu}\varphi_j-\frac{1}{2}\varphi_i\text{M}^{ij}\varphi_j\right],
\end{equation}
where $\varphi_i$ is a set of fields and $\text{M}$, $\text{N}^{\mu}$ are symmetric and antisymmetic matrices, respectively, that depend on the metric. Then the gravitational one-loop counterterms are given by
\begin{equation}\label{thooftdiv}
    S_{\text{ct}}(g)=-\frac{\mu^{-\varepsilon}}{(4\pi)^2 \varepsilon}\int \sqrt{-g} \ \text{tr}\left[\frac{1}{12}\text{Y}_{\mu\nu}\text{Y}^{\mu\nu}+\frac{1}{2}\left(\text{M}-\text{N}^{\mu}\text{N}_{\mu}+\frac{\mathds{1}}{6}R\right)^2+\frac{\mathds{1}}{120}C^2\right],
\end{equation}
where the trace is understood over the $i,j$ indices, $\mathds{1}\equiv \delta_{ij}$ and
\begin{equation}
\text{Y}_{\mu\nu}=\partial_{\mu}\text{N}_{\nu}-\partial_{\nu}\text{N}_{\mu}+\text{N}_{\mu}\text{N}_{\nu}-\text{N}_{\nu}\text{N}_{\mu}.
\end{equation}
Setting $\lambda_{s}=0$ and integrating by parts, the action \eqref{hsgeneral} can be written as
\begin{equation}\label{hslambda0}
    S_s(\varphi,g)=\frac{(-1)^{s+1}}{2}\int \sqrt{-g}\left[\varphi_i P^{ij}\nabla^2\varphi_j+\varphi_i\text{X}^{ij}\varphi_j\right],
    \end{equation}
where $P$ and X are symmetric matrices and we use the short notations $\varphi_i=\varphi_{\mu_1\dots\mu_s}$, $P^{ij}=P^{\mu_1\dots\mu_s,\nu_1\dots\nu_s}$ and $\text{X}^{ij}=\text{X}^{\mu_1\dots\mu_s,\nu_1\dots\nu_s}$.
To reduce \eqref{hslambda0} into the form \eqref{thooftact} we proceed as follows. First we double the fields $\varphi_i$ by complexifying them, then we perform the redefinition $\varphi^*_i\rightarrow \varphi^*_j(P^{-1})^{j}_{\ i}$ and obtain the action
\begin{equation}\label{complexact}
S'_s(\varphi,g)=(-1)^{s+1}\int \sqrt{-g}\left[\varphi_i^*\delta^{ij}\nabla^2\varphi_j+\varphi_i^*(P^{-1})^i_{\ k}\text{X}^{kj}\varphi_j\right],
\end{equation}
where $\delta^{ij}=\eta^{\mu\nu}$ for $s=1$, $\delta^{ij}=\frac{1}{2}\left(\eta^{\mu\rho}\eta^{\nu\sigma}+\eta^{\mu\sigma}\eta^{\nu\rho}-\frac{1}{2}\eta^{\mu\nu}\eta^{\rho\sigma}\right)$ for $s=2$ and so on. Note that the last redefinition breaks general covariance, which is recovered in final result. Expanding the kinetic term we find 
\begin{equation}
  \varphi_i^*\delta^{ij}\nabla^2\varphi_j=\varphi_i^*\delta^{ij}g^{\mu\nu}\partial_{\mu}\partial_{\nu}\varphi_j +\varphi_i^*(\text{N}^{\mu})^{ij}\partial_{\mu}\varphi_j+\varphi_i^*J^{ij}\varphi_j,
\end{equation}
then the action~\eqref{complexact} is of the form~\eqref{thooftact} with $\text{M}=J+P^{-1}\text{X}$. After taking into account the right factors due to the doubling, we go back to real fields and for the particular case of \eqref{hslambda0}, we have $J=\text{N}^{\mu}\text{N}_{\mu}$, $\text{X}=m_s^2 P$ and $\text{M}=\text{N}^{\mu}\text{N}_{\mu}+m_s^2\mathds{1}$. Moreover, the matrix $\text{N}^{\mu}$ corresponds to the generalized spin connection and the quantity $\text{Y}_{\mu\nu}$ coincides with the generalized curvature
\begin{equation}
    \mathcal{R}_{\mu\nu}\equiv\left[\nabla_{\mu},\nabla_{\nu}\right]=\Sigma_{\ a}^{b}R^a_{ \ b\mu\nu},
\end{equation}
where $R^{a}_{ \ b\mu\nu}$ is the Riemann tensor and $\Sigma_{\ a}^{b}$ are the generators of the Lorentz group.

The counterterms \eqref{thooftdiv} reduce to 
\begin{equation}\label{hsdiv}
    S_{s}^{\text{ct}}(g)=-\frac{\mu^{-\varepsilon}}{(4\pi)^2 \varepsilon}\int \sqrt{-g}\left\{ \frac{1}{12}\text{tr}\mathcal{R}_{\mu\nu}\mathcal{R}^{\mu\nu}+\text{tr}\mathds{1}\left[\frac{1}{2}\left(m_s^2+\frac{1}{6}R\right)^2+\frac{1}{120}C^2\right]\right\},
\end{equation}
where $\text{tr}\mathds{1}=\text{d}_s=(s+1)^2$ is the dimension of the representation. The computation of $\text{tr}\mathcal{R}_{\mu\nu}\mathcal{R}^{\mu\nu}$ amounts to compute of the product of two generators traced on the indices of the representation. Using formulas of \autoref{lorentz}, we get
\begin{align}
\int\sqrt{-g} \ \text{tr}\mathcal{R}_{\mu\nu}\mathcal{R}^{\mu\nu}&=-\frac{\text{d}_s(\text{d}_s-1)}{12}\int\sqrt{-g}R_{\mu\nu\rho\sigma}R^{\mu\nu\rho\sigma}\\
&=-\frac{\text{d}_s(\text{d}_s-1)}{6}\int\sqrt{-g}\left(C^2+\frac{1}{6}R^2\right).
\end{align}
The final expression for the gravitational counterterms generated by the action \eqref{hslambda0} is 
\begin{equation}\label{hsdivfin}
    S_{s}^{\text{ct}}(g)=-\frac{\mu^{-\varepsilon}}{(4\pi)^2 \varepsilon}\int \sqrt{-g} \ \text{d}_s\left[\frac{m_s^2}{6}R+\frac{m^4_s}{2}+\frac{8-5\text{d}_s }{360}C^2+\frac{7-\text{d}_s}{432}R^2\right].
\end{equation}
We have explicitly checked this formula for $s=1,2,3$ by extracting the tensor $\text{N}^{\mu}$ for all the cases. Moreover, the coefficients of $C^2$ and $R^2$ are in agreement with~\cite{Christensen:1978md}. The nonminimal terms~\eqref{bnm} can be included in the analysis by means of the substitution $\text{X}\rightarrow\text{X}+\text{B}_s$ in~\eqref{hslambda0}.

\section{Fermions}
\label{sec:fermions}
\setcounter{equation}{0}
It is possible to treat fermionic higher-spin multiplets along the same line of the previous section. We consider spinor multiplets $\psi_{\mu_1\ldots\mu_s}$, completely symmetric and traceless in the spacetime indices and satisfying the condition
\begin{equation}\label{gammapsi}
\gamma^{\mu}\psi_{\mu\mu_2\ldots\mu_s}=0.
\end{equation}
In flat spacetime, the most general action that we can build is 
\begin{equation}\label{fermionflat}
    S_{s}(\psi)=(-1)^s\int\bar{\psi}_{\mu_1\ldots\mu_s}\left(i\gamma^{\rho}\partial_{\rho}-m_{s}\right)\psi^{\mu_1\ldots\mu_s}.
\end{equation}
The field $\psi_{\mu_1\ldots\mu_s}$ transforms according to the irreducible representation  $\left(\frac{s+1}{2},\frac{s}{2}\right)\oplus \left(\frac{s}{2},\frac{s+1}{2}\right)$ of the Lorentz group and contains $2(s+2)(s+1)$ degrees of freedom. Note that there is no analogue of $\lambda_s$, as well as there are no self interactions compatible with renormalizability. In this respect, the coupling of fermionic higher-spin multiplets to gravity is more simple then the bosonic case, where renormalizable self interactions and the running of their couplings should be taken into account for a complete analysis.

The action (\ref{fermionflat}) coupled to gravity is 
\begin{equation}\label{fermioncurved}
    S_{s}(\psi,e)=(-1)^s\int e \ \bar{\psi}_{\mu_1\ldots\mu_s}g^{\mu_1\nu_1}\dots g^{\mu_s\nu_s}\left(i\gamma^{a}e^{\rho}_a\nabla_{\rho}-m_{s}\right)\psi_{\nu_1\ldots\nu_s},
\end{equation}
where $e\equiv \text{det}\left(e^{\rho}_a\right)$, $e^{\rho}_a$ is the vierbein and the covariant derivative $\nabla$ acts on the fermionic field as
\begin{equation}
    \nabla_{\rho}\psi_{\mu_1\ldots\mu_s}=\partial_{\rho}\psi_{\mu_1\ldots\mu_s}-\frac{1}{8}\omega_{\ b\rho}^{a}\left[\gamma^b,\gamma_a\right]\psi_{\mu_1\ldots\mu_s}-\Gamma_{\rho\mu_1}^{\alpha}\psi_{\alpha\ldots\mu_s}-\ldots-\Gamma_{\rho\mu_s}^{\alpha}\psi_{\mu_1\ldots\alpha}
\end{equation}
\begin{equation}
    \omega_{\ b\rho}^{a}\equiv e^{a}_{\nu}\left(\partial_{\rho}e^{\nu}_b+\Gamma_{\alpha\rho}^{\nu}e^{\alpha}_b\right), \qquad e^{\mu}_ae^{b}_{\mu}=\delta_a^b.
\end{equation}
The quantity $\omega_{\ b\rho}^{a}$ is called ``spin connection".

\subsection[Counterterms]{Counterterms for any spin $s+1/2$} \label{countertermfermion}
Using the same notation  of \autoref{counterterms0} we write the action (\ref{fermioncurved}) as
\begin{equation}\label{fermioncompact}
    S_{s+1/2}(e,\psi)=(-1)^s\int e\left[i\bar{\psi}_iP^{ij}\slashed{\nabla}\psi_j-m_{s}\bar{\psi}_iP^{ij}\psi_j\right], \qquad \slashed{\nabla}=\gamma^ae_a^{\mu}\nabla_{\mu},
\end{equation}
where now in the indices $i,j\ldots$ collect also the fermionic ones. 

In order to obtain the gravitational counterterms generated by the fermionic multiplets we proceed as follows. First we perform the redefinition
\begin{equation}\label{fermred}
    \psi_j\rightarrow i\slashed{\nabla}\chi_j+m_{s}\chi_j, \qquad \bar{\psi}_i\rightarrow\bar{\psi}_i,
\end{equation}
where $\chi_{\mu_1\ldots\mu_s}$ can be completely symmetric, $\gamma$-tracelss and transverse, without loss of generality, obtaining the action
\begin{equation}
    S_{s+1/2}(e,\bar{\psi},\chi)=(-1)^{s+1}\int e \left[\bar{\psi}_iP^{ij}\slashed{\nabla}\slashed{\nabla}\chi_j+m_{s}^2\bar{\psi}_iP^{ij}\chi_j\right].
\end{equation}
Then, using the anticommutation rules of the Dirac matrices, we write
\begin{equation}
    \slashed{\nabla}\slashed{\nabla}\chi=g^{\mu\nu}\nabla_{\mu}\nabla_{\nu}\chi+\frac{1}{2}\gamma^{\mu}\gamma^{\nu}
\mathcal{R}_{\mu\nu}\chi,
\end{equation}
Finally the action (\ref{fermioncompact}) can be written as
\begin{equation}
    S_{s+1/2}(g,\bar{\psi},\chi)=(-1)^{s+1}\int\sqrt{-g}\left[\bar{\psi}_iP^{ij}g^{\mu\nu}\nabla_{\mu}\nabla_{\nu}\chi_j+\bar{\psi}_iP^{ij}\big(m_s^2+\frac{1}{2}\gamma^{\mu}\gamma^{\nu}
\mathcal{R}_{\mu\nu}\big)_j^{\ l}\chi_l\right],
\end{equation}
which is of the form (\ref{complexact}) with $X=P\left(m_s^2+\frac{1}{2}\gamma^{\mu}\gamma^{\nu}
\mathcal{R}_{\mu\nu}\right)$. Finally, following the steps of \autoref{counterterms0}, we obtain the following expression for the counterterms
\begin{equation}
    S_{s}^{\text{ct}}(g)=\frac{\mu^{-\varepsilon}}{(4\pi)^2 \varepsilon}\int \sqrt{-g}\left\{ \frac{1}{12}\text{tr}\mathcal{R}_{\mu\nu}\mathcal{R}^{\mu\nu}+\text{tr}\left[\frac{1}{2}\left(\mathds{1}m_s^2+\frac{\mathds{1}}{6}R+\frac{1}{2}\gamma^{\mu}\gamma^{\nu}
\mathcal{R}_{\mu\nu}\right)^2+\frac{\mathds{1}}{120}C^2\right]\right\},
\end{equation}
where we have included a minus one for the fermionic loop and a factor $1/2$ for the doubling caused by the field redefinition (\ref{fermred}).

In order to derive the counterterms for fermions in the representation $\left(\frac{s+1}{2},\frac{s}{2}\right)\oplus\left(\frac{s}{2},\frac{s+1}{2}\right)$ we proceed as explained in \autoref{lorentz}. The result is
\begin{equation}\label{hsdivferm}
    S_{s}^{\text{ct}}(g)=-\frac{\mu^{-\varepsilon}}{(4\pi)^2 \varepsilon}\int \sqrt{-g} \ \text{d}_s\left[\frac{m_s^2}{12}R-\frac{ m^4_s}{2}+\frac{49-10\text{d}_s }{720}C^2+\frac{\text{d}_s-4}{432}R^2\right].
\end{equation}

\section{Asymptotic freedom}
\label{AF}
\setcounter{equation}{0}
In this section we determine whether it is possible to obtain asymptotic freedom in both the gravitational couplings $\alpha$ and $\xi$. For this purpose we need to study the full system of the renormalization group equations and then look for solutions with asymptotic behavior that gives a well-defined free theory in the UV. Since the gravitational sector of the theory contains higher derivatives we require that a two-derivative term survive in the UV and the cosmological term be negligible. This means that it is necessary to fix also the asymptotic behavior of $\zeta$ and $\Lambda$, which would then require to know the behavior of $m_s$ for every $s$ present in the theory, since they contribute to the beta functions of $\zeta$ and $\Lambda$. More precisely, after rescaling the metric fluctuation $h_{\mu\nu}\rightarrow \sqrt{\alpha}h_{\mu\nu}$, the asymptotic Lagrangian can be written (omitting the indices) as  
\begin{equation}
    \kappa^2 \mathcal{L}_{UV}\sim\frac{1}{2}h\Delta_4h-\frac{\alpha}{6\xi}h\Delta'_4h+\alpha\zeta h\Delta_2 h+2\Lambda(\sqrt{\alpha}h+\alpha h^2),
\end{equation}
where $\Delta_{4}(\Delta'_{4})$ and $\Delta_2$ are four-derivative and two-derivative operators, respectively. In order to have that both the higher-derivative terms be dominant in the UV, we look for solutions such that $\alpha/\xi$ flows to a constant. Moreover, for what explained above, $\alpha\zeta$ should flow to a constant while $\Lambda\sqrt{\alpha}$ should be negligible. For the purpose of this paper we restrict the analysis only to the beta functions of $\alpha$ and $\xi$, since unwanted behavior for $\zeta$ and $\Lambda$ can still be compensated by the power counting. In fact, since the cosmological term has no derivatives it can always be neglected in the UV, compared to the linearized Einstein term. The latter can also be considered as ``small" compared to the higher-derivative ones. A full study in some minimal cases pointed out below is postponed to a future publication. 

From~\eqref{hsdivfin} and~\eqref{hsdivferm} we can extract the UV contributions to the beta functions
\begin{equation}\label{eq:bosonbeta}
    \Delta\alpha_s^{(b)}=\frac{\text{d}_s(5\text{d}_s-8)}{180},\qquad \Delta\xi_s^{(b)}=\frac{\text{d}_s(7-\text{d}_s)}{72},
\end{equation}
\begin{equation}
    \Delta\alpha_s^{(f)}=\frac{\text{d}_s(10\text{d}_s-49)}{360},\qquad \Delta\xi_s^{(f)}=\frac{\text{d}_s(\text{d}_s-4)}{72},
\end{equation}
where the superscript $(b)$ and $(f)$ denotes the contributions of bosons and fermions, respectively, and 
\begin{equation}
    \text{d}_s=\begin{cases}
      (s+1)^2, \qquad \text{bosons}\\[2ex]
      2(s+1)(s+2),\qquad \text{fermions}.
    \end{cases}
\end{equation}
The contributions to the beta functions are 
\begin{equation}
    \Delta x_{\text{tot}}=\Delta x+\sum_{s=1}^{s'}\left(N_s\Delta x_s^{(b)}+N_{s+\frac{1}{2}}\Delta x_s^{(f)}\right),\qquad  x=\alpha,\xi,
\end{equation}
where $s'$ is some positive integer number and $\Delta\alpha$, $\Delta\xi$ are given by~\eqref{deltas}. We anticipate that we have found no solutions with $N_{s+\frac{1}{2}}\neq 0$ for any $s$. Therefore, in what follows we directly drop the contributions of fermions. 

The two beta functions read
\begin{equation}\label{eq:betas}
    \beta_{\alpha}=\frac{\text{d}\alpha}{\text{d}t}=\Delta\alpha_{\text{tot}}\alpha^2, \qquad \beta_{\xi}=\frac{\text{d}\xi}{\text{d}t}=\Delta\xi_{\text{tot}}\xi^2.
\end{equation}
The first equation can be solved straightforwardly since $\Delta\alpha_{\text{tot}}$ is constant. Fixing the initial condition $\alpha(0)=\alpha_0$ the solution is
\begin{equation}\label{alphasol}
    \alpha(t)=\frac{\alpha_0}{1-\alpha_0\Delta\alpha_{\text{tot}}t}.
\end{equation}
\begin{table}[t]
\centering
\begin{tabular}{|c|c|c|c|c|c|c|c|c|c|c|c|c|c|c|c|c|c|c|c|c|c|c|c|} 
\hline
& \multicolumn{22}{|c|}{Solutions}\\
\hline
$N_1$ & 0 & 0 & 0 & 0 & 0 & 0 & 0 & 0 & 0 & 1 & 1 & 1 & 1 & 1 & 1 & 1 & 1 & 2 & 2 & 2 & 2 & 2 \\
$N_2$ & 0 & 0 & 1 & 2 & 3 & 4 & 5 & 6 & 7 & 0 & 0 & 1 & 2 & 3 & 5 & 6 & 7 & 0 & 1 & 2 & 3 & 5 \\
 $N_3$ & 1 & 2 & 1 & 1 & 1 & 0 & 0 & 0 & 0 & 1 & 2 & 1 & 1 & 1 & 0 & 0 & 0 & 1 & 1 & 1 & 1 & 0 \\
$N_{s>3}$ & 0 & 0 & 0 & 0 & 0 & 0 & 0 & 0 & 0 & 0 & 0 &0&0&0&0&0&0&0&0&0&0&0 \\
\hline
\end{tabular}

\vspace{0.2cm}

\centering
\begin{tabular}{|c|c|c|c|c|c|c|c|c|c|c|c|c|c|c|c|c|c|c|c|c|c|c|c|} 
\hline
$N_1$ & 2 & 3 & 3 & 3 & 3 & 3 & 4 & 4 & 4 & 4 & 5 & 5 & 5 & 5 & 6 & 6 & 6 & 7 & 7 & 8 & 8 & 9 \\
$N_2$ & 6 & 0 & 1 & 2 & 3 & 6 & 0 & 1 & 2 & 3 & 0 & 1 & 2 & 3 & 0 & 1 & 2 & 1 & 2 & 1 & 2 & 2 \\
$N_3$ & 0 & 1 & 1 & 1 & 1 & 0 & 1 & 1 & 1 & 1 & 1 & 1 & 1 & 1 & 1 & 1 & 1 & 1 & 1 & 1 & 1 & 1 \\
$N_{s>3}$ & 0 & 0 & 0 & 0 & 0 & 0 & 0 & 0 & 0 & 0 & 0 &0&0&0&0&0&0&0&0&0&0&0 \\
\hline
\end{tabular}
\caption{\emph{The solutions of the system (\ref{sys0}).}}
\label{tabsol0}
\end{table}
However, the second equation is more involved because it depends on the ratio $\alpha/\xi$. Since we are interested the UV behavior, we insert the asymptotic expression $\alpha(t)\sim -1/(\Delta\alpha_{\text{tot}}t)$ of~\eqref{alphasol} in the second equation of~\eqref{eq:betas} and look for solutions that behave like $\xi(t)\sim C_2/t$ in the UV. The result is a second order equation in $C_2$. Choosing the smallest solution for $C_2$ we find
\begin{equation}
    \frac{\alpha}{\xi}\sim \frac{\Delta\alpha_{\text{tot}}  (5+6 \overline{\Delta \xi})}{3 (\Delta\alpha_{\text{tot}}-5 ) \Delta\alpha_{\text{tot}} +\sqrt{\Delta\alpha_{\text{tot}} ^2 \left[175-60 \overline{\Delta \xi}+9
   (\Delta\alpha_{\text{tot}}-10 ) \Delta\alpha_{\text{tot}} \right]}},
\end{equation}
\begin{equation}
  \Delta\xi_{\text{tot}}^{\text{UV}}\sim -\frac{3}{2}\Delta\alpha_{\text{tot}}+\frac{3 }{10}\Delta\alpha_{\text{tot}} ^2-\frac{1}{10} \sqrt{\Delta\alpha_{\text{tot}} ^2 \left[175-60 \overline{\Delta \xi}+9 (\Delta\alpha_{\text{tot}}-10
   ) \Delta\alpha_{\text{tot}}\right]},
\end{equation}
where $\overline{\Delta\xi}=\Delta\xi_{\text{tot}}-\Delta\xi$.
We need to solve the system\footnote{In general we need to add to the system the no-tachyon condition for every $s$, which is always satisfied when $\lambda_s=0$.}
\begin{equation}\label{sys0}
    \begin{cases}
      \Delta\alpha_{\text{tot}}\left(N_s\right)<0\\[2ex]
      \Delta\xi_{\text{tot}}^{\text{UV}}\left(\Delta\alpha_{\text{tot}},\overline{\Delta\xi},N_s\right)<0.
    \end{cases}
    \end{equation}
In~\autoref{tabsol0} we collect the solutions of~\eqref{sys0} in terms of $N_s$. We find 44 different solutions. Note the first solution in the top part of~\autoref{tabsol0}, where only one irreducible rank-3 field is enough to obtain asymptotic freedom. Moreover, for $s>3$ every $N_s$ vanishes. This can be explained as follows. All the contributions to $\Delta \alpha_s^{(b)}$ for $s>0$ have positive sign. Therefore, $N_{s>0}$ cannot be too large, otherwise $\Delta \alpha_{\text{tot}}>0$. On the other hand, for $\Delta \xi_{s}^{(b)}$ all the contributions with $s>1$ are negative. Since the $N_s$ are the same for both $\Delta \alpha_s^{(b)}$ and $\Delta \xi_{s}^{(b)}$, there are only some values of $s$ such that there exist values of $N_s$ to solve the system. The inclusion of $N_0$ scalar fields, as well as the contribution of standard model, enlarges the set of solutions. For example, it is possible to have scalar fields up to $N_0=37$. We do not present those solutions in~\autoref{tabsol0} since they do not add more to the analysis. However, in this respect, it is worth to note that from~\eqref{eq:bosonbeta} the contributions of $s=0$ and $s=2$ have opposite signs for both $\alpha$ and $\xi$
\begin{align}
    N_0\Delta\alpha_0^{(b)}+N_2\Delta\alpha_2^{(b)}&= -\frac{1}{60}N_0+\frac{37}{20}N_2,\\
    N_0\Delta\xi_0^{(b)}+N_2\Delta\xi_2^{(b)}&= \ \frac{1}{12}N_0-\frac{1}{4}N_2.
\end{align}
This opens the possibility to have a fixed point \`a la Banks-Zacs~\cite{Banks:1981nn}, which requires further investigation\footnote{D. Anselmi and M.Piva, in preparation.}.

Finally, we recall that the no-tachyon condition on cosmological spacetime of ref.~\cite{Anselmi:2020lpp} should also be included. In the notation of this paper, the condition reads
\begin{equation}\label{eq:notachcosmo}
    \frac{\alpha}{\xi}>\frac{1}{16}.
\end{equation}
However, in~\cite{Anselmi:2020lpp} the bound~\eqref{eq:notachcosmo} is obtained by neglecting the cosmological constant directly from the action. Therefore, we do not include the cosmological no-tachyon condition, since its general form with nonvanishing $\Lambda$ is not known yet. Nevertheless, we have checked that, in the hypothetical case where the cosmological constant can be neglected, the inclusion of~\eqref{eq:notachcosmo} reduces the set of solutions of~\autoref{tabsol0} only by a relatively small amount, from 44 to 29. Most of the interesting minimal solutions are still present, including the case $(N_1,N_2,N_3)=(0,0,1)$. 

In the next section, we generalize these results to $\lambda_s\neq 0$ for $s=1,2$.

\section{Bosons with nonminmal kinetic term}
\label{examples}
\setcounter{equation}{0}
The procedure used in Section~\ref{counterterms0} works only for $\lambda_s=0$, i.e. for actions with minimal kinetic terms, where the derivatives are contracted between each other. In this section we compute the counterterms with generic $\lambda_s$ for $s=1,2$ using Feynman diagrams. The computation for $s>2$ is involved and require more sophisticated techniques \cite{Barvinsky:1985an} or generalization of the algorithm \cite{tHooft:1974toh}, such as \cite{Pronin:1996rv}.

\begin{figure}
\center
   \includegraphics[width=12cm]{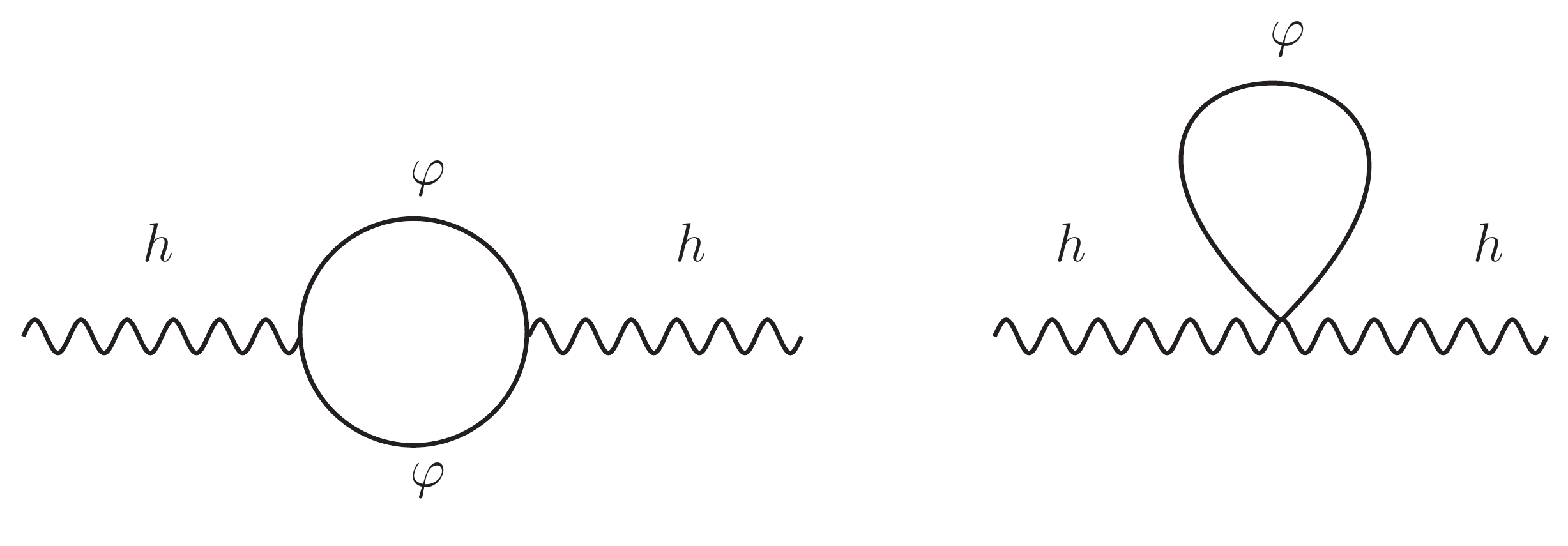}
   \caption{\emph{The Feynman diagrams that contribute to the graviton self energy. The wavy lines denote the graviton field, the solid lines denote the irreducible rank-$s$ matter multiplet.}}
   \label{diagrams}
\end{figure}

In order to derive the counterterms using Feynman diagrams, we expand around flat spacetime $\eta_{\mu\nu}=\text{diag}\left(1,-1,-1,-1\right)$ and write
\begin{equation}
    g_{\mu\nu}=\eta_{\mu\nu}+2\kappa h_{\mu\nu},
\end{equation}
where $h_{\mu\nu}$ is the graviton field. Then the gravitational divergent terms can be obtained by computing the contributions to the graviton self energy given by the sum of diagrams in~\autoref{diagrams}. Therefore, we need only the three-leg vertex $\varphi-\varphi-h$ and the four-leg vertex $\varphi-\varphi-h-h$.

\subsection{Rank 1}
The action for $s=1$ and $\lambda_1\neq 0$ in curved spacetime is
\begin{equation}\label{spin1curved}
    S_1(\varphi)=-\frac{1}{2}\int\sqrt{-g}\left(\nabla_{\mu}\varphi_{\nu}\nabla^{\mu}\varphi^{\nu}+\lambda_1\nabla_{\mu}\varphi_{\nu}\nabla^{\nu}\varphi^{\mu}-m_1^2\varphi_{\mu}\varphi^{\mu}\right).
\end{equation}
Expanding around flat spacetime, the propagator in momentum space reads
\begin{equation}
    D_{\mu\nu}(p)=\frac{-i}{p^2-m_1^2+i\epsilon}\left(\eta_{\mu\nu}-\frac{p_{\mu}p_{\nu}}{m_1^2}\right)+\left.\frac{-i}{p^2-\frac{m_1^2}{1+\lambda_1}}\right|_{\text{f}}\frac{p_{\mu}p_{\nu}}{m_1^2},
\end{equation}
where we have already chosen the appropriate quantization prescriptions in order to preserve unitarity. In particular, the subscript ``f" denotes that, in the amplitudes, the branch cuts associated with that pole are treated by means of formula~\eqref{av}. The propagator has poles in $p^2=m_1^2$ and $p^2=m_{10}^2\equiv m_1^2/(1+\lambda_1)$, which have positive and negative residues, respectively. To see this we proceed as follows. First we choose the rest frame $p_{\mu}=\left(m_1,0,0,0\right)$ and write the propagator as a $4\times4$ matrix $\mathbb{D}_1$. Then, we compute the residue matrix for each pole
\begin{equation}
    -i(p^2-m_1^2)\mathbb{D}_1\left.\right|_{p^2=m_1^2}=\left(\setlength\arraycolsep{5pt}\begin{array}{cccc}
0 & 0 & 0 & 0\\
0 & 1 & 0 & 0\\
0 & 0 & 1 & 0\\
0 & 0 & 0 & 1\\
\end{array}\right), \quad
  -i(p^2-m_0^2)\mathbb{D}_1\left.\right|_{p^2=m_0^2}=\left(\setlength\arraycolsep{5pt}\begin{array}{cccc}
\frac{-1}{1+\lambda_1} & 0 & 0 & 0\\
0 & 0 & 0 & 0\\
0 & 0 &0 & 0\\
0 & 0 & 0 & 0\\
\end{array}\right).
\end{equation}
In order to establish the spin of the submultiplet we look at the number of the nonzero residues, which corresponds to the dimension of the submultiplet. For a given $s$ the dimension of a submultiplet of spin $l$ is $2l+1$. The pole in $m_1^2$ has three nonzero residues, while the pole in $m_0^2$ has one. Therefore, we have $l=1$ and $l=0$, respectively. For $s>1$ we need an additional step where we turn the propagator, which is a rank-$2s$ tensor, into a $\text{d}_s\times \text{d}_s$ matrix.  

After extracting the vertices with a Mathematica program, the diagrams of Figure~\ref{diagrams} give the following contributions to the beta functions
\begin{equation}\label{eq:deltalambda1}
    \Delta\alpha_1=\frac{32+10\lambda_1-35\lambda_1^2}{120},\qquad \Delta\xi_1=\frac{4+8\lambda_1+17\lambda_1^2}{24}.
\end{equation}
\begin{equation}
    \Delta\zeta_1=-\frac{8+12\lambda_1+9\lambda_1^2}{12(1+\lambda_1)}m_1^2,\qquad \Delta\Lambda_{1}=-\frac{4+6\lambda_1+3\lambda_1^2}{2(1+\lambda_1)^2}m_1^4.
\end{equation}
Note that the special value $\lambda_1=-1$ is singular. The reason is that this choice corresponds to Proca theory, which is nonrenormalizable. Since it not possible to obtain a nonrenormalizable theory as a limit of a renormalizable one, the correct result in this case is obtained by imposing $\lambda_1=-1$ directly in the action~\eqref{spin1curved} and then perform the calculations. 

The system~\eqref{sys0} with $N_{s>1}=0$ and the contributions~\eqref{eq:deltalambda1} only, together with the no-tachyon condition $\lambda_1>-1$, has no solution. We conclude that the rank-1 field theory~\eqref{spin1curved} alone is not enough for asymptotic freedom.

\subsection{Symmetric rank 2}
The action for $s=2$ and $\lambda_2\neq 0$ in curved spacetime is
\begin{equation}\label{spin2curved}
    S_2(\varphi)=\frac{1}{2}\int\sqrt{-g}\left(\nabla_{\mu}\varphi_{\nu\rho}\nabla^{\mu}\varphi^{\nu\rho}+\lambda_2\nabla_{\mu}\varphi_{\nu\rho}\nabla^{\nu}\varphi^{\mu\rho}-m_2^2\varphi_{\mu\nu}\varphi^{\mu\nu}\right).
\end{equation}
The propagator around flat spacetime reads
\begin{equation}
    D_{\mu\nu\rho\sigma}(p)=\frac{\Pi^{(2)}_{\mu\nu\rho\sigma}}{p^2-m^2_2+i\epsilon}-\left.\frac{2\Pi^{(1)}_{\mu\nu\rho\sigma}}{(2+\lambda_2)p^2-2 m_2^2}\right|_{\text{f}}+\frac{\tilde{\Pi}^{(0)}_{\mu\nu\rho\sigma}}{(4+3\lambda_2)-4m_2^2+i\epsilon},
\end{equation}
where $\tilde{\Pi}^{(0)}\equiv \Pi^{(0)}+3\bar{\Pi}^{(0)}-\bar{\bar{\Pi}}^{(0)}$ and the $\Pi^{(i)}$ are the spin-2 projectors defined in Appendix~\ref{projectors}. The action (\ref{spin2curved}) describes a massive spin-2, spin-1 and spin-0 fields with squared masses 
\begin{equation}
   m_2^2, \qquad m_{21}^2\equiv\frac{2 m_2^2}{2+\lambda_2}, \qquad m_{20}^2\equiv\frac{4m_2^2}{4+3\lambda_2},
\end{equation}
respectively. The no-tachyon condition in this case is $\lambda_2>-4/3$. The signs in front of the projectors show that the spin-1 field has a negative residue in the propagator and must be quantized as fakeon. The signs of the residues and the dimensions of the submultiplets can be derived by applying the same method explained in the previous subsection. We write the propagator as a $9\times 9$ matrix by using $\varphi_{00}, \varphi_{01}, \varphi_{02},$ etc... as basis. Then the diagonalized matrices of the residues at the poles in $m_2^2$, $m_{21}^2$ and $m_{20}^2$ are
\begin{equation}\label{eq:dofrank2}
     \left(\setlength\arraycolsep{5pt}\begin{array}{cc}
\mathds{1}_{5} &  \\
 & \mymathbb{0}_4 \\
\end{array}\right), \quad 
   \frac{2}{2+\lambda_2}\left(\setlength\arraycolsep{3pt}\begin{array}{ccc}
\mymathbb{0}_{5} & &  \\
 & -\mathds{1}_{3} & \\
 & & 0\\
\end{array}\right),
\quad
\frac{4}{4+3\lambda_2}\left(\setlength\arraycolsep{5pt}\begin{array}{cc}
\mymathbb{0}_{8} &  \\
 & 1 \\
\end{array}\right),
\end{equation}
respectively. We stress that no additional degree of freedom can be turned on by the coupling to a general background since~\eqref{eq:dofrank2} already describes all the possible degrees of freedom of $\varphi_{\mu\nu}$ ($\mathbf{9}=\mathbf{2}\oplus\mathbf{1}\oplus\mathbf{0}$). In principle we could have considered also a general action of a reducible multiplet with nonvanishing trace. Besides~\eqref{eq:dofrank2} (with more involved dependence of masses and residues on the parameters) there would be an additional scalar degree of freedom even around flat spacetime, which occupies the ``slot" of the would-be Boulware-Deser ghost. For this reason, also reducible multiplets can be coupled to gravity without any inconsistency.

In curved spacetime, the traceless condition for the multiplet $\varphi_{\mu\nu}$ reads $g^{\mu\nu}\varphi_{\mu\nu}=0$. In order to proceed with the computations using Feynman diagrams is convenient to switch to a field variable $\tilde{\varphi}_{\mu\nu}$ such that $\eta^{\mu\nu}\tilde{\varphi}_{\mu\nu}=0$. Then, in the new variable, the propagator is the same, while the three-leg vertex gets contributions from the quadratic terms in the old variable, the four-leg vertex gets contributions from the three-leg vertex in the old variable and so on. For the computations of this section is enough to write $\varphi(\tilde{\varphi})$ up to the order $\kappa^2$
\begin{equation}
    \varphi_{\mu\nu}=\tilde{\varphi}_{\mu\nu}+\frac{1}{4}\eta_{\mu\nu}\tilde{\varphi}_{\rho\sigma}\left[2\kappa h^{\rho\sigma}+\kappa^2\left( hh^{\rho\sigma}-h^{\rho\alpha}h_{\alpha}^{\sigma}\right)\right]+\mathcal{O}(\kappa^3).
\end{equation}
The result of the computation gives the following contributions to the beta functions
\begin{equation}
    \Delta\alpha_2=\frac{28416+76160 \lambda _2+66672 \lambda _2^2+10320 \lambda _2^3-16904 \lambda _2^4-10260 \lambda _2^5-1915 \lambda
   _2^6}{240 \left(2+\lambda _2\right){}^2 \left(4+3 \lambda _2\right){}^2},
   \end{equation}
   \begin{equation}
   \Delta\xi_2=\frac{-192+160 \lambda _2+2052 \lambda _2^2+3636 \lambda _2^3+2951 \lambda _2^4+1230 \lambda _2^5+220 \lambda _2^6}{12
   \left(2+\lambda _2\right){}^2 \left(4+3 \lambda _2\right){}^2},
\end{equation}
\begin{equation}
    \Delta\zeta_2=-\frac{\left(192+624 \lambda _2+800 \lambda _2^2+524 \lambda _2^3+185 \lambda _2^4+30 \lambda _2^5\right)}{2
   \left(2+\lambda _2\right){}^2 \left(4+3 \lambda _2\right){}^2}m_2^2,
    \end{equation}
    \begin{equation}
    \Delta\Lambda_{2}=-3\frac{192+384 \lambda _2+288 \lambda _2^2+100 \lambda _2^3+15 \lambda _2^4}{2\left(2+\lambda _2\right)^2 \left(4+3 \lambda _2\right)^2}m_2^4.
\end{equation}
Again, we see that there are singularities for the special values $\lambda_2=-2,-4/3$. In flat spacetime, the former corresponds to the traceless version of Pauli-Fierz theory, while the latter corresponds to a conformal invariant kinetic term\footnote{Conformal invariance is broken by the presence of the mass term.}. Both cases are nonrenormalizable and cannot be viewed as limit of the renormalizable ones, in analogy with the rank-1 case.
\begin{figure}[t]
    \centering
    \includegraphics[width=8cm]{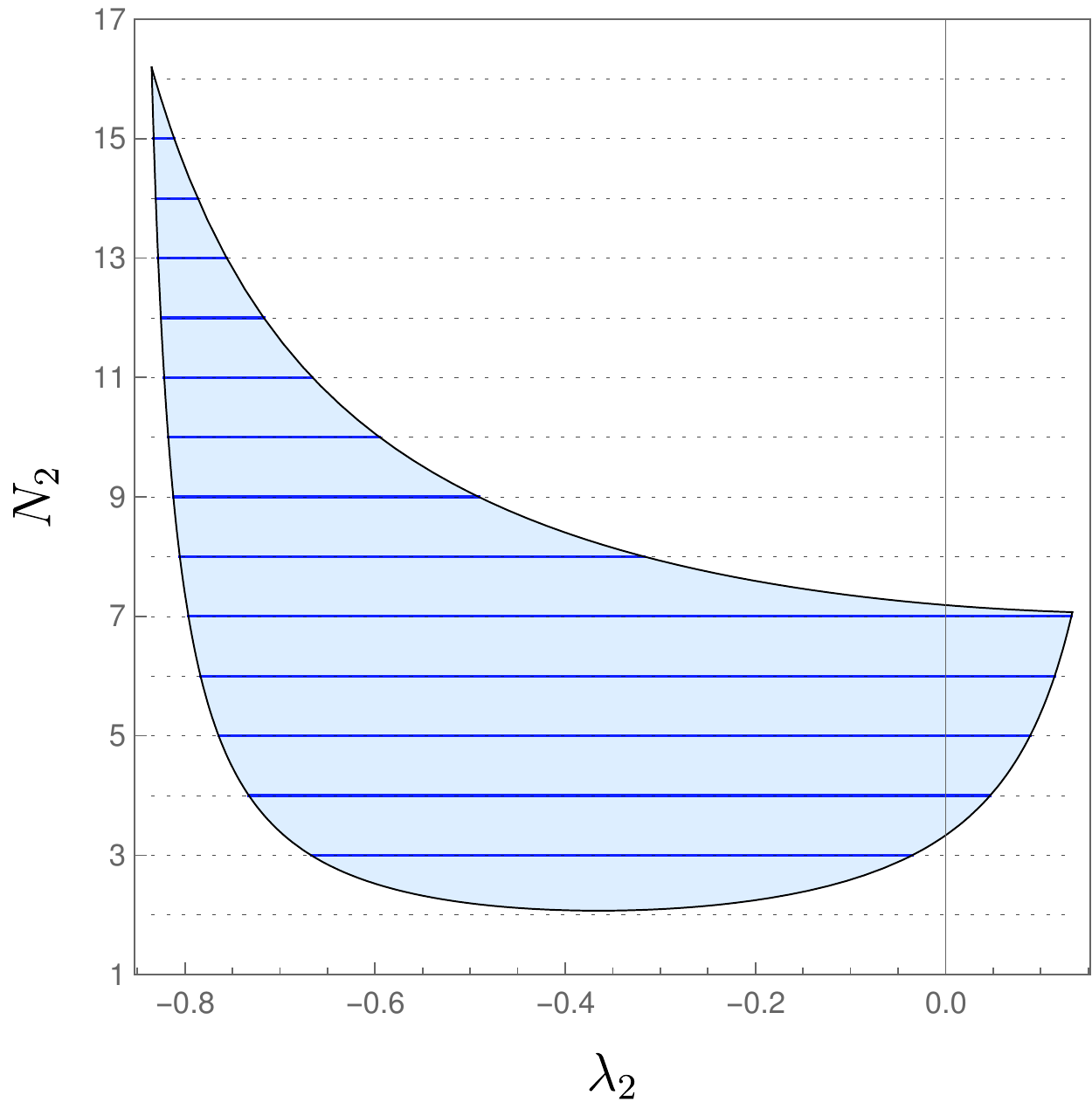} \includegraphics[width=8cm]{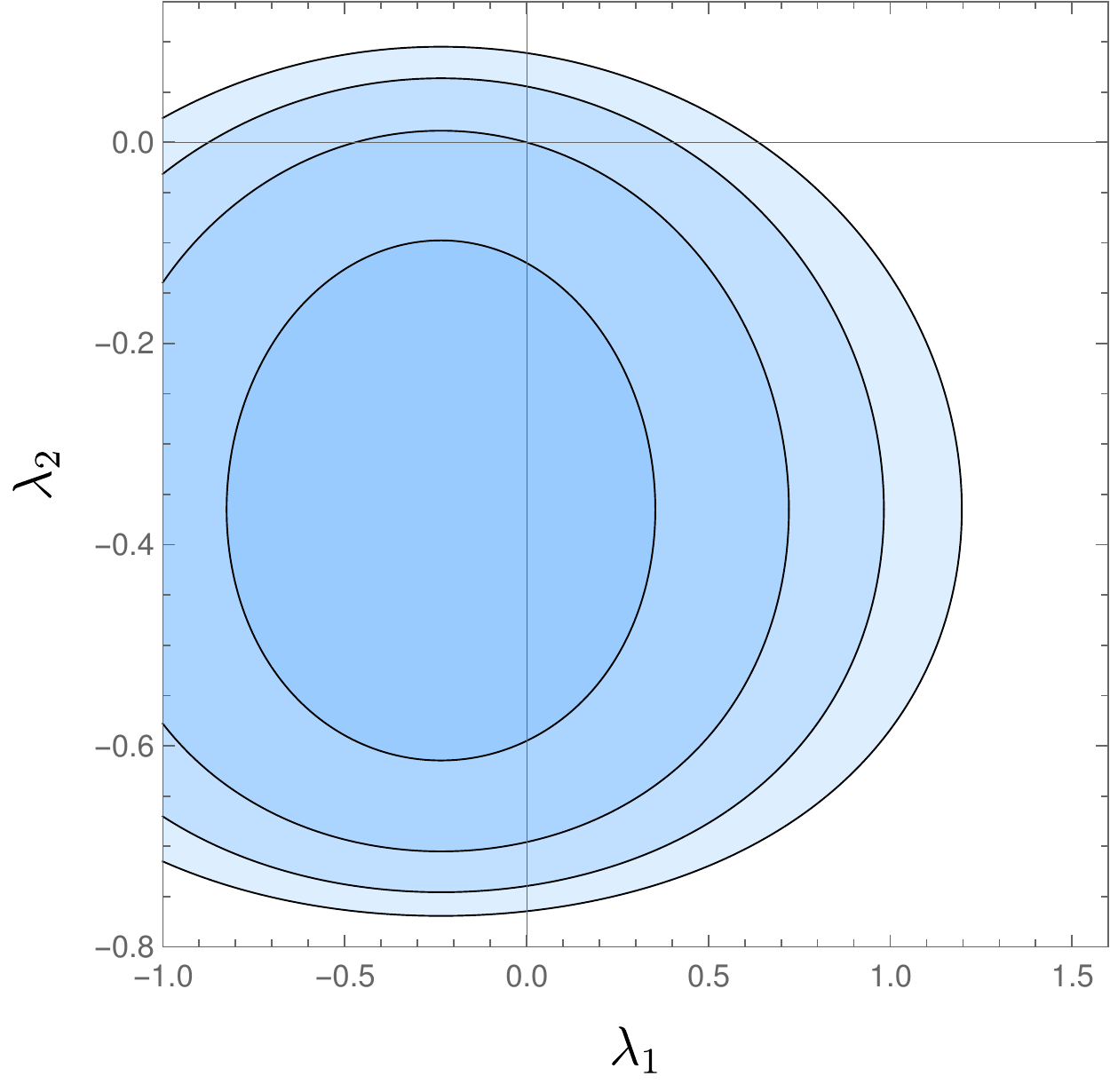}
    \caption{\emph{In the left panel, the values of $N_2$ and the ranges of $\lambda_2$ (solid blue lines inside the shaded region) compatible with asymptotic freedom with $N_1=0$. The vertical line $\lambda_2=0$ identifies the solutions of~\autoref{tabsol0} with $N_3=0$. In the right panel, the regions in the plane $(\lambda_1,\lambda_2)$ compatible with asymptotic freedom with $N_1=1$ and $N_2=3,4,5,6$.}}
    \label{fig:solN}
\end{figure}
Solving the system \eqref{sys0} in this case, together with the contributions from rank-1 tensor and the no-tachyon conditions $\lambda_1>-1$ and $\lambda_2>-4/3$, we find that the set of solutions is enlarged compared to that of~\autoref{AF} (with $s=1,2$ only). In particular, having nonvanishing $\lambda_{1,2}$ produces new solutions with higher $N_1$ and $N_2$. As an example, in the left panel of~\autoref{fig:solN} we show the result for $N_1=0$ and general $N_2$. In this case we find
\begin{equation}\label{l2range}
    -0.83494...<\lambda_2<0.13331..., 
\end{equation}  
depending on $N_2$, which is bounded by two rational functions
\begin{equation}\label{N2range}
   \frac{\left(2+\lambda _2\right)^2 \left(4+3 \lambda _2\right)^2}{P(\lambda_2)}<N_2<\frac{\left(2+\lambda _2\right)^2 \left(4+3 \lambda _2\right)^2}{Q(\lambda_2)},
\end{equation}
where $P$ and $Q$ are polynomials of degree 6 such that $P>Q$ for $\lambda_2$ in the range~\eqref{l2range}. From~\eqref{N2range}, the possible values for $N_2$ in the range~\eqref{l2range} are 
\begin{equation}
3\leq N_2\leq 17.    
\end{equation}
In the right panel of~\autoref{fig:solN} we show the allowed ranges for $(\lambda_1, \lambda_2)$ fixing $N_1=1$ and varying $N_2$ between 3 and 6. Increasing $N_2$ enlarges the possible values for $(\lambda_1, \lambda_2)$. We conclude that the choice $\lambda_1=\lambda_2=0$ is not a special one. The analysis in the case of higher $s$ is more involved and goes beyond the purpose of this paper.

\section{Metric affine quantum gravity}
\label{sec:mag}
A class of theories that contains the higher-spin multiplets studied in this paper is metric affine gravity~\cite{WeylTSM, Cartan:1923zea} (see also~\cite{Hehl:1994ue} for a review). Indeed, a theory of a metric tensor $g_{\mu\nu}$ and an affine connection $\Upsilon^{\rho}_{\ \mu\nu}$ can be viewed as the theory~\eqref{lhd} coupled to a rank-3 tensor field, which contains several irreducible multiplets. The most general renormalizable action invariant under diffeomorphisms contains all the possible terms (up to dimension 4) obtained by contracting the Riemann tensor, the torsion and the non-metricity\footnote{We recall that when either torsion or non-metricty are nonvanishing, the Riemann tensor has less symmetries.} (see for example~\cite{Percacci:2019hxn} for a recent analysis). Then the affine connection can be written as 
\begin{equation}
    \Upsilon_{ \ \mu\nu}^{\rho}=\Gamma_{ \ \mu\nu}^{\rho}+\Omega_{ \ \mu\nu}^{\rho},
\end{equation}
where $\Gamma_{ \ \mu\nu}^{\rho}$ is the Levi-Civita connection. The quantity $\Omega_{ \ \mu\nu}^{\rho}$ is a rank-3 tensor, being the difference of two connections, which contains 64 degrees of freedom that can be decomposed as 
\begin{equation}
    \mathbf{64}=\mathbf{3}\oplus\mathbf{2}^5\oplus\mathbf{1}^9\oplus \mathbf{0}^5.
\end{equation}
\begin{table}
\centering
\begingroup
\setlength{\tabcolsep}{10pt} 
\renewcommand{\arraystretch}{1.5}
\begin{tabular}{|c|c|c|c|c|c| } 

\hline
Fields & \multicolumn{4}{|c|}{$l$ \ $m_{3l}^2$ \ $\sigma$}\\
\hline

$\tilde{S}_{\mu\nu\rho}$ & \ 3 \  \ $m^2$ \  $+$ & \ 2 \ \ $\frac{3 m^2}{3+\lambda}$ \  \ $-$ \ & \ 1 \  \ $\frac{9 m^2}{9+5\lambda}$\  \ $+$ \ & \ 0 \ \ $\frac{3 m^2}{3+2\lambda}$\  \ $-$\  \\
\hline
$A_{\mu\nu\rho}$ & \ 0 \  \ $m^2$ \ \ $+$ \  & \ 1 \  \ $\frac{3 m^2}{3+\lambda}$ \  \ $-$ \  &  & \\
\hline
$\tilde{H}^{(s)}_{\mu\nu\rho}$ & \ 2 \ \ $m^2$ \  \ $+$ \ & \ 1 \ \ $\frac{2 m^2}{2+\lambda-\lambda'}$ \ \ $-$ \ & \ 2 \ \ $\frac{6 m^2}{6+\lambda+\lambda'}$ \ \ $-$ \   & \ 1 \ \ $\frac{9 m^2}{9+5\lambda-4\lambda'}$ \  \ $+$ \ \\
\hline
$\tilde{H}^{(a)}_{\mu\nu\rho}$ & \ 2 \  \ $m^2$ \  \ $+$ \ & \ 1 \ \ $\frac{2 m^2}{2+\lambda+\lambda'}$\  \ $-$ \ & \ 2 \  \ $\frac{6 m^2}{6+\lambda-\lambda'}$ \ \ $-$ \  & \ 1 \ \ $\frac{3 m^2}{3+5\lambda}$ \  \ $+$ \ \\
\hline
$\text{Tr}$ &  \ 1 \ \ $m^2$ \ \ $+$ \  & \ 0 \  \ $\frac{ m^2}{1+\lambda}$\  \ $-$ \ &  & \\
\hline
\end{tabular}
\endgroup
\caption{\emph{The spin ($l$), squared mass ($m_{3l}^2$) and sign ($\sigma$) of the residue of the degrees of freedom contained in $\Omega_{\mu\nu\rho}$. The parameters $m$, $\lambda$ and $\lambda'$ must be considered different in every row. Tilde denotes the traceless part of the fields. ``{\rm Tr}" collectively denotes the three traces.}}
\label{rank3tab}
\end{table}
A rank-$3$ tensor field $\Omega_{\mu\nu\rho}$ can be split into 4 reducible components 
\begin{equation}\label{eq:sahh}
    \Omega_{\mu\nu\rho}=S_{\mu\nu\rho}+A_{\mu\nu\rho}+H_{\mu\nu\rho}^{(s)}+H_{\mu\nu\rho}^{(a)},
\end{equation}
where $S_{\mu\nu\rho}$ and $A_{\mu\nu\rho}$ are are completely symmetric and antisymmetric, respectively, while $H^{(s)}_{\mu\nu\rho}$ and $H^{(a)}_{\mu\nu\rho}$ are symmetric and antisymmetric under the exchange of $\mu$ and $\nu$, respectively\footnote{We could choose, instead, (anti)symmetry under the exchange between $\nu$ and $\rho$. Both choices give the same contributions to the beta functions in the case considered below.}. Each of them can be further decomposed into irreducible components. In particular, the traceless part of $S_{\mu\nu\rho}$ corresponds to the field $\varphi_{\mu\nu\rho}$, which is the only one that contains a spin-3 particle. The field $A_{\mu\nu\rho}$ can be described by a rank-1 field $\varphi_{\mu}$, as well as the traces $S^{\rho}_{\rho\mu}$, $H^{(s)\rho}_{\mu\rho}$ and $H^{(a)\rho}_{\mu\rho}$. The traceless parts of the fields $H^{(s)}_{\mu\nu\rho}$ and $H^{(a)}_{\mu\nu\rho}$, instead, do not belong to the class described by the action~\eqref{hsgeneral}. Indeed, their general quadratic action in flat spacetime is~\cite{Anselmi:2020opi}
\begin{equation}\label{hookflat}
    S_3^{(s/a)}=\frac{-1}{2}\int\left(\partial_{\sigma}\tilde{H}_{\mu\nu\rho}\partial^{\sigma}\tilde{H}^{\mu\nu\rho}+\lambda\partial^{\sigma}\tilde{H}_{\sigma\mu\nu}\partial_{\rho}\tilde{H}^{\rho\mu\nu}+\lambda'\partial^{\sigma}\tilde{H}_{\sigma\mu\nu}\partial_{\rho}\tilde{H}^{\rho\nu\mu}-m^2\tilde{H}_{\mu\nu\rho}\tilde{H}^{\mu\nu\rho}\right),
\end{equation}
where we have generically denoted both fields with $\tilde{H}_{\mu\nu\rho}$. The propagating degrees of freedom described by each component of $\Omega_{\mu\nu\rho}$ are collected in~\autoref{rank3tab}. Several fields have negative residue and can be quantized with the fakeon prescription.

If we minimally couple~\eqref{hookflat} to gravity, set $\lambda^{s,a}=\lambda'^{s,a}=0$ and use the algorithm of~\autoref{counterterms0}, we find the following contributions to the beta functions
\begin{equation}\label{eq:betahook}
    \Delta\alpha_3^{(s,a)}=\frac{56}{15}, \qquad \Delta\xi_3^{(s,a)}=-\frac{2}{3}, \qquad \Delta\zeta_3^{(s,a)}=\Delta\zeta_3, \qquad \Delta\Lambda_3^{(s,a)}=\Delta\Lambda_3. 
\end{equation}
Taking into account the decomposition~\eqref{eq:sahh}, we look for solutions to the system~\eqref{sys0} with
\begin{equation}
\Delta x_{\text{tot}}=\Delta x+N_1\Delta x_1+\left(N_3^{(s)}+N_3^{(a)}\right)\Delta x_3^{(s,a)}+N_3\Delta x_3, \qquad x=\alpha,\xi,
\end{equation}
$$
 N_1\in\{0,1,2,3,4\}, \qquad N_3,N_3^{(s)},N_3^{(a)}\in \{0,1\}.
$$
Excluding the solutions already known from~\autoref{tabsol0}, we find asymptotic freedom for the values collected in~\autoref{tab:solhsha}.

Note that, with our symmetry choice for the indices of $H^{(s)}_{\mu\nu\rho}$ and $H^{(a)}_{\mu\nu\rho}$, the solutions with $N_3^{(a)}=0$ and $N_1\leq 2$ contain torsionless theories with nonvanishing non-metricity as a particular case. Instead, choosing  anti/symmetry under the exchange of $\nu$ and $\rho$ for $H^{(a/s)}_{\mu\nu\rho}$ explicitly shows that metric-compatible theories with nonvanishing torsion ($N_3=N_3^{(s)}=0$ and $N_1\leq 2$) are not among the solutions. Again, we expect that the inclusion of all nonminimal kinetic terms in the analysis enlarges the set of solutions. Therefore, quantum metric affine theories of gravity might be asymptotically free.

\begin{table}[t]

\centering
\begin{tabular}{|c|c|c|c|c|c|c|c|c|c|c|c|}
\hline
& \multicolumn{11}{|c|}{Solutions}\\
\hline
 $N_1$& 0 & 0 & 0 & 1 & 1 & 2 & 2 & 3 & 3 & 4 & 4 \\
 $N_3$& 0 & 1 & 1 & 1 & 1 & 1 & 1 & 1 & 1 & 1 & 1 \\
 $N_3^{(s)}$& 1 & 0 & 1 & 0 & 1 & 0 & 1 & 0 & 1 & 0 & 1 \\
 $N_3^{(a)}$&1 & 1 & 0 & 1 & 0 & 1 & 0 & 1 & 0 & 1 & 0 \\
 \hline
\end{tabular}
\caption{\emph{Solutions of the system~\eqref{sys0} in the case of a rank-$3$ tensor $\Omega_{\mu\nu\rho}$ with minimal and non-mixed kinetic terms.}}
\label{tab:solhsha}
\end{table}

In light of the results of this section, it would be interesting to perform a detailed analysis of this class of theories. Although the most general action contains several parameters, some simplifications can be adopted. For example, if we consider the theory of a rank-3 tensor where the mixed kinetic terms and odd interactions are absent, they would not be generated by renormalization, even if we include the coupling to quantum gravity. In fact, this corresponds to a $\mathbb{Z}_2$ symmetry for $\Omega_{\mu\nu\rho}$ and only 4-field interactions with even number of the same fields can be generated.  

Finally, quantum metric affine theories of gravity can contain a propagating physical spin-3 particle. This provides a useful arena for exploring the possibility that particles of spin higher than 2 might exist in nature (see~\cite{Marzo:2021esg} for a recent work on propagation in flat spacetime of spin-$3$ particles in the free-field limit).

\section{Conclusions}
\label{sec:concl}
\setcounter{equation}{0}
We have shown that the renormalizable and unitary theory of quantum gravity can be made asymptotically free by adding suitable matter multiplets. The new matter multiplets contain both standard and purely virtual degrees of freedom. We have computed the contributions to the gravitational beta functions of fields with arbitrary spin, both bosonic and fermionic, with minimal kinetic term in the case of bosons. We have performed the analysis of the relevant RG equations and established the number of fields necessary to obtain asymptotic freedom in the gravitational couplings. We have found several solutions (with bosonic fields only), including some minimal ones that deserve more attention for future work. We have extended the analysis including nonminimal kinetic terms for rank-1 and symmetric rank-2 fields. The results show that the set of solutions compatible with asymptotic freedom is enlarged. Finally, this work opens to  interesting applications in the case of metric affine theories of gravity, suggesting that there might be room for asymptotic freedom. 

\subsection*{Acknowledgments}
We thank D. Anselmi, S. Giaccari, C. Marzo, A. Melis and A. Tseytlin for useful discussions and comments. This work is supported by the Estonian Research Council grant MOBTT86 and by the EU through the European Regional Development Fund CoE program TK133 ``The Dark Side of the Universe".

\appendix
\renewcommand{\thesection}{\Alph{section}} \renewcommand{\theequation}{%
\thesection.\arabic{equation}} \setcounter{section}{0}

\section{Lorentz algebra formulas}\label{lorentz}
\setcounter{equation}{0}
The Lorentz group has generators $\Sigma^{ab}$ such that
\begin{equation}
    \Sigma^{ab}=-\Sigma^{ba}, \qquad \text{tr}\Sigma^{ab}=0, 
\end{equation}
\begin{equation}
    \left[\Sigma^{ab},\Sigma^{cd}\right]=\frac{1}{2}\left(\Sigma^{ad}\eta^{cb}-\Sigma^{bd}\eta^{ac}+\Sigma^{ac}\eta^{bd}-\Sigma^{bc}\eta^{ad}\right).
\end{equation}
Consider an irreducible representation $r=\left(A,B\right)$ of the Lorentz group. Then we have   
\begin{equation}
    \text{tr}\left(\Sigma^{ab}\Sigma^{cd}\right)=C(r)\frac{\eta^{ad}\eta^{bc}-\eta^{ac}\eta^{bd}}{2}, \qquad \Sigma^{ab}\Sigma_{ba}=C_{2}(r)\mathds{1}_{r},
\end{equation}
where $C(r)$ is called ``Dynkin index", $C_2(r)$ are the eigenvalues of the quadratic Casimir operator and $\mathds{1}_r$ is the identity in the representation $r$. Note that both $C(r)$ and $C_2(r)$ depend on the representation. The two can be related to each other by taking the trace of the Casimir operator finding
\begin{equation}
    C(r)=\frac{C_2(r)}{6}\text{d}_s.
\end{equation}
In order to derive the coefficient $C_2(r)$ for any representation $(A,B)$ we remind that the algebra of the (homogeneous) Lorentz group is isomorphic to the algebra of $SU(2)\otimes SU(2)$. Labelling $A^a$ and $B^a$ the generators of the two $SU(2)$ algebras we have 
\begin{equation}
    \text{Tr}\mathds{1}_A=2A+1, \qquad A^aA^a=C_2^A\mathds{1}_A, \qquad C^A_2=A(A+1),
\end{equation}
and analogous formulas for $B^a$. Then for a representation $(A,B)$ of the Lorentz group we have
\begin{equation}
    \Sigma^{ab}\Sigma_{ba}=A^aA^a\otimes\mathds{1}_B+\mathds{1}_A\otimes B^aB^a=(C^A_2+C^B_2)\mathds{1}_A\otimes \mathds{1}_B.
\end{equation}
Therefore, the Dynkin index reads
\begin{equation}\label{eq:dynkin}
    C(r)=\frac{(2A+1)(2B+1)}{6}\left[A(A+1)+B(B+1)\right].
\end{equation}
Using~\eqref{eq:dynkin} we can extract the coefficient for the particular cases we are interested in.

For $r=\left(\frac{s}{2},\frac{s}{2}\right)$ we have $A=B=\frac{\sqrt{\text{d}_s}-1}{2}$, thus
\begin{equation}\label{cbosons}
 C_2(r)=\frac{\text{d}_s-1}{2} \qquad \Rightarrow   \qquad  C(r)=\frac{\text{d}_s(\text{d}_s-1)}{12}.
\end{equation}

For $r=\left(\frac{1}{2},0\right)\oplus\left(0,\frac{1}{2}\right)$ we have 
\begin{equation}\label{fermiongen}
\Sigma_f^{ab}=-\frac{1}{8}\left[\gamma^a,\gamma^b\right], \qquad \Sigma_f^{ab}\Sigma_f^{ba}=\frac{3}{4}\mathds{1}_r
\end{equation}
and therefore
\begin{equation}
  C_2(r)=\frac{3}{4}, \qquad   C(r)=\frac{1}{2}.
\end{equation}

For $r=\left(\frac{s}{2},\frac{s}{2}\right)\otimes\left[\left(\frac{1}{2},0\right)\oplus\left(0,\frac{1}{2}\right)\right]$ we have the tensor product of two representations. In general, the generators $\text{T}(r)$ in a representation $r=r_1\otimes r_2$ are given by
\begin{equation}
   \text{T}_r=\text{T}_{r_1}\otimes\mathds{1}_{r_2}+\mathds{1}_{r_1}\otimes\text{T}_{r_2},
\end{equation}
where $\text{T}_{r_i}$ and $\mathds{1}_{r_i}$ are generators and the identity in the representation $r_i$, respectively.
In this case we have
\begin{equation}
    \Sigma^{ab}=\Sigma_{\text{B}}^{ab}\otimes\mathds{1}_{f}+\mathds{1}_{\text{B}}\otimes\Sigma^{ab}_f,
\end{equation}
where $\Sigma_{\text{B}}^{ab}$ are the generators in the representation $(\frac{s}{2},\frac{s}{2})$ and $\Sigma_f^{ab}$ are those in \eqref{fermiongen}. The trace of two generators is
\begin{align}
     \text{tr}\left(\Sigma^{ab}\Sigma^{cd}\right)&=4 \text{tr}\left(\Sigma_{\text{B}}^{ab}\Sigma_{\text{B}}^{cd}\right)+\text{d}_{\text{B}}\text{tr}\left(\Sigma_{f}^{ab}\Sigma_{f}^{cd}\right)=\left[4\frac{\text{d}_{\text{B}}(\text{d}_{\text{B}}-1)}{12}+\frac{\text{d}_{\text{B}}}{2}\right]\frac{\eta^{ad}\eta^{bc}-\eta^{ac}\eta^{bd}}{2}
\end{align}
and therefore
\begin{equation}\label{cproduct}
    C(r)=\frac{\text{d}_{\text{B}}(2\text{d}_{\text{B}}+1)}{6},
\end{equation}
where $\text{d}_{\text{B}}=(s+1)^2$.

Finally, in order to obtain the coefficient $C(r)$ for the spinor representation $r=\left(\frac{s+1}{2},\frac{s}{2}\right)\oplus\left(\frac{s}{2},\frac{s+1}{2}\right)$, used in~\autoref{countertermfermion}, we need to subtract from (\ref{cproduct}) the contribution of  $\left(\frac{s-1}{2},\frac{s}{2}\right)\oplus\left(\frac{s}{2},\frac{s-1}{2}\right)$. This can be done by subtracting the contribution of $\left(\frac{s-1}{2},\frac{s-1}{2}\right)\otimes\left[\left(\frac{1}{2},0\right)\oplus\left(0,\frac{1}{2}\right)\right]$, from which we need to subtract the contribution of $\left(\frac{s-2}{2},\frac{s-2}{2}\right)\otimes\left[\left(\frac{1}{2},0\right)\oplus\left(0,\frac{1}{2}\right)\right]$ and so on. Labeling (\ref{cproduct}) as $\mathcal{C}(\text{d}_{\text{B}})$, this is obtained by means of the formula
\begin{equation}
    C(r)=\sum_{j=0}^{\sqrt{\text{d}_{\text{B}}}-1}(-1)^j\mathcal{C}\left[(\sqrt{\text{d}_{\text{B}}}-j)^2\right]=\frac{1}{12} \left(-\sqrt{\text{d}_{\text{B}}}+\text{d}_{\text{B}}+4 
    \text{d}_{\text{B}}^{3/2}+2\text{d}_{\text{B}}^2\right),
\end{equation}
which in terms of the dimension of the representation $\text{d}_s$ turns into the more simple form 
\begin{equation}\label{cproductirr}
    C(r)=\frac{\text{d}_s(\text{d}_s-1)}{24}.
\end{equation}
Note that this last expression as a function of $\text{d}_s$ is just one half of (\ref{cbosons}).

\section{Spin-2 projectors}\label{projectors}
\setcounter{equation}{0}
Starting from the transverse and longitudinal projectors for vectors
\begin{eqnarray}
\theta_{\mu\nu}&\equiv & \eta_{\mu\nu}-\frac{p_{\mu}p_{\nu}}{p^2},\\
\omega_{\mu\nu}&\equiv & \frac{p_{\mu}p_{\nu}}{p^2},
\end{eqnarray}
we define the spin-2 projectors as
\begin{eqnarray}
\Pi^{(2)}_{\mu\nu\rho\sigma}&\equiv&\frac{1}{2}(\theta_{\mu\rho}\theta_{\nu\sigma}+\theta_{\mu\sigma}\theta_{\nu\rho})-\frac{1}{3}\theta_{\mu\nu}\theta_{\rho\sigma},\\
\Pi^{(1)}_{\mu\nu\rho\sigma}&\equiv&\frac{1}{2}(\theta_{\mu\rho}\omega_{\nu\sigma}+\theta_{\mu\sigma}\omega_{\nu\rho}+\theta_{\nu\rho}\omega_{\mu\sigma}+\theta_{\nu\sigma}\omega_{\mu\rho}),\\
\Pi^{(0)}_{\mu\nu\rho\sigma}&\equiv&\frac{1}{3}\theta_{\mu\nu}\theta_{\rho\sigma},\\
\bar{\Pi}^{(0)}_{\mu\nu\rho\sigma}&\equiv&\omega_{\mu\nu}\omega_{\rho\sigma},\\
\bar{\bar{\Pi}}^{(0)}_{\mu\nu\rho\sigma}&\equiv&\theta_{\mu\nu}\omega_{\rho\sigma}+\theta_{\rho\sigma}\omega_{\mu\nu}.
\end{eqnarray}

\bibliographystyle{JHEP} 
\bibliography{mybiblio}

\end{document}